# Solution of the nonrelativistic wave equation in the tridiagonal representation approach

A. D. Alhaidari

*Saudi Center for Theoretical Physics, P.O. Box 32741, Jeddah 21438, Saudi Arabia*

**Abstract**: We choose a complete set of square integrable functions as basis for the expansion of the wavefunction in configuration space such that the matrix representation of the nonrelativistic time-independent wave operator is tridiagonal and symmetric. Consequently, the matrix wave equation becomes a symmetric three-term recursion relation for the expansion coefficients of the wavefunction in this basis. The recursion relation is then solved exactly in terms of orthogonal polynomials in the energy. Some of these polynomials are not found in the mathematics literature. The asymptotics of these polynomials give the phase shift of the continuous energy scattering states and the spectrum for the discrete energy bound states. Depending on the space and boundary conditions, the basis functions are written in terms of either the Laguerre or Jacobi polynomials. The tridiagonal requirement limits the number of potential functions that yield exact solutions of the wave equation. Nonetheless, the class of exactly solvable problems in this approach is larger than the conventional class (see Table 12). We also give very accurate results for cases where the wave operator matrix is not tridiagonal but its elements could be evaluated either exactly or numerically with high precision.

We are honored to dedicate this work to Prof. Hashim A. Yamani on the occasion of his 70th birthday.

**PACS**: 03.65.Pm, 03.65.Ge, 03.65.Fd, 03.65.Ca
**Keywords**: tridiagonal representation, orthogonal polynomials, recursion relation, energy spectrum, phase shift, generalized potentials.

# 1. Introduction

Exact solutions of the wave equation result in a better understanding of the physical system modeled by the equation. Such understanding cannot be achieved by approximate or numerical solutions especially in situations such as phase transitions, critical coupling, limiting cases, singularities, etc. To obtain an exact solution of the time-independent wave equation, one has to provide the wavefunction for all energies (continuous and discrete) and give the energy spectrum of the discrete bound states (if they exist) and the phase shift of the continuum scattering states. There are many methods of solution of the non-relativistic time-independent wave equation. These include, but not limited to, shape invariance, supersymmetry, factorization, algebraic methods, operator algebra, point canonical transformations, etc. Here, we will not give an overview of these methods but the interested reader can consult the vast literature on the subject including many of the popular books on quantum mechanics.

In the "*Tridiagonal Representation Approach* (TRA)", we expand the wavefunction over a complete set of square integrable functions $\{\phi_n(x)\}$ in configuration space. That is, we write $|\psi(E,x)\rangle = \sum_n f_n(E)|\phi_n(x)\rangle$, where the expansion coefficients $\{f_n(E)\}$ are some appropriate functions of the energy. They contain all physical information (both structural and dynamical) about the system. The basis functions $\{\phi_n(x)\}$, on the other hand, carry very little information about the specific system under study. They contain only kinematic information shared by a whole class of systems that have the same expansion structure. Such kinematic information is limited to, say, the angular momentum quantum number and a length scale. Now, with this expansion, the wave equation $H|\psi\rangle = E|\psi\rangle$ becomes $\sum_n f_n H|\phi_n\rangle = E\sum_n f_n|\phi_n\rangle$. Projecting from left by $\langle\phi_m|$, we obtain the equation $\sum_n f_n\langle\phi_m|H|\phi_n\rangle = E\sum_n f_n\langle\phi_m|\phi_n\rangle$, which is a generalized eigenvalue matrix wave equation

$$\sum_n H_{m,n} f_n = E\sum_n \Omega_{m,n} f_n \,, \tag{1}$$

where $\Omega_{n,m} = \langle\phi_n|\phi_m\rangle$ is the overlap matrix of the basis elements (i.e., matrix representation of the identity). For a reason that will shortly be apparent, we require that the matrix representation of the wave operator $\mathcal{J}_{nm} = H_{nm} - E\,\Omega_{nm}$ be tridiagonal and symmetric. This will be possible only for special potential functions. If so, the wave equation (1) turns into a symmetric three-term recursion relation for the expansion coefficients $\{f_n(E)\}$. Writing $f_n(E) = f_0(E)P_n(E)$ makes $\{P_n(E)\}$ a set of orthogonal polynomials satisfying the three-term recursion relation with the seed $P_0(E) = 1$. Thus, the wave function becomes $|\psi(E,x)\rangle = f_0(E)\sum_n P_n(E)|\phi_n(x)\rangle$, where $f_0(E)$ is an overall normalization factor whose square is proportional to the positive weight function associated with the orthogonal polynomials $\{P_n(E)\}$ [1]. With all polynomials $\{P_n(E)\}$ and weight function $f_0^2(E)$ being identified, the wavefunction $\psi(E,x)$ is now exactly realized and according to the postulates of quantum mechanics, the associated physical system is well defined. For example, the energy spectrum of the bound states (if they exist) is given by the spectrum formula of the polynomial $P_n(E)$. Moreover, the scattering phase shift of the continuum energy states (if they exist) is given by the asymptotics ($n\to\infty$) of these same polynomials [1-3]. The general form of the asymptotics is given as Eq. (46) in section 3 below where we also show how to extract from it the phase shift and energy spectrum. Hence, the central ingredient in the TRA is this set of orthogonal energy polynomials.

The solutions presented here below are either for problems in one dimension or in three dimensions with spherical symmetry. Aside from the Coulomb and isotropic oscillator, all other 3D problems are solved for *S*-wave (zero angular momentum). Additionally, it is worth noting that the TRA could also be used successfully to handle 3D non-central separable problems as illustrated in Ref. [4].

On the other hand, if the matrix representation of the wave operator $\mathcal{J}$ is not tridiagonal but its elements could either be derived exactly or be evaluated with high precision, we can still obtain very accurate solutions. For these cases, we choose a representation in which the Hamiltonian matrix is still tridiagonal but not $\Omega$ (i.e., the basis elements are neither orthogonal nor tri-thogonal).

## 2. Solution in the Laguerre basis

Let $x$ be the configuration space coordinate and $y(x)$ be a transformation to a dimensionless coordinate system such that $y \geq 0$. We choose the following square integrable functions as a complete set of basis in the new space with coordinate $y$

$$\phi_n(x) = \mathcal{A}_n y^\alpha e^{-\beta y} L_n^\nu(y), \tag{2}$$

where $L_n^\nu(y)$ is the Laguerre polynomial of degree $n$ in $y$ and $\mathcal{A}_n = \sqrt{\frac{\Gamma(n+1)}{\Gamma(n+\nu+1)}}$. The parameters $\{\alpha, \beta, \nu\}$ are real and dimensionless such that $\beta > 0$ and $\nu > -1$. In Appendix A, we show that the matrix representation of the wave operator $\mathcal{J} = -\frac{1}{2}\frac{d^2}{dx^2} + V(x) - E$ in this basis is tridiagonal and symmetric if and only if $\frac{dy}{dx} = \lambda y^a e^{by}$, $2\beta = 1 + b$ and $V(x) - E = y^{2a} e^{2by}\left(U_0 + U_1 y^{-1} + U_2 y^{-2}\right)$ where the set $\{U_i\}$ are real parameters. The real dimensionless constants $a$ and $b$ are fixed for each choice of configuration space associated with a given problem and such that $b > -1$. Moreover, the length scale parameter $\lambda$ has an inverse length dimension. Additionally, we should select one of two alternatives scenarios:

(1) $2\alpha = 1 + \nu - a$ and $U_2 = \frac{\lambda^2}{8}\left[\nu^2 - (1-a)^2\right]$, or (3a)

(2) $2\alpha = 2 + \nu - a$ and $U_0 = \frac{\lambda^2}{8}\left(1 - b^2\right)$. (3b)

Now, the overlap matrix of the basis elements $\Omega_{n,m} = \langle \phi_n | \phi_m \rangle$ reads as follows:

$$\Omega_{n,m} = \mathcal{A}_n \mathcal{A}_m \int_0^\infty y^\nu e^{-y} L_n^\nu(y) L_m^\nu(y) \left(y^{2\alpha-\nu-a} e^{-2by}\right) dy \equiv \langle n | y^{2\alpha-\nu-a} e^{-2by} | m \rangle, \tag{4}$$

where we have defined $\langle y | n \rangle = \mathcal{A}_n y^{\nu/2} e^{-y/2} L_n^\nu(y)$. For the two cases above, this becomes a tridiagonal matrix only if:

(3a) $b = 0$ and $a = \frac{1}{2}$ or $a = 0$.

(3b) $b = 0$ and $a = \frac{1}{2}$ or $a = 1$.

However, if $\Omega$ is not tridiagonal, then the non-tridiagonal component $-E\,\Omega$ of the wave operator must be eliminated by a counter term in the Hamiltonian. This is necessary to maintain the tridiagonal matrix structure of the wave operator $H - E\,\Omega$ so that the wave equation becomes a three-term recursion relation to be solved exactly for the expansion coefficients of the wavefunction in terms of orthogonal polynomials in the energy. Now, since the potential function is energy independent then the counter term that cancels $-E\,\Omega$ must come from the kinetic energy part of the Hamiltonian. For case (3a), this means that the balancing term must come form the $U_2$ term of the expression $V(x) - E$ whereas for the (3b) case it comes from the $U_0$ term. Hence, non-tridiagonal $\Omega$ is allowed in the above two scenarios only as follows:

(3a) $b = 0$ and $a = 1$ with $V_{const} - E = \lambda^2 \nu^2 / 8$, and

(3b) $b = 0$ and $a = 0$ with $V_{const} - E = \lambda^2 / 8$,

where $V_{const}$ is the constant part of the potential. In fact, we can obtain the same allowed set of values of the configuration space parameters $a$ and $b$ by noting that in the above equations for $V(x)-E$, the energy independence of the potential dictates that $b=0$ and $a$ should assume one of three values $a=\left\{0,\frac{1}{2},1\right\}$.

In the process of solving the matrix wave equation (1), we need to evaluate matrix elements of the form $\langle n|f(y)|m\rangle$, where $f(y)$ is some given function. This could be done exactly only for special cases (e.g., $f=1$, $f=y$, $f=y^2$). However, in general we must use any appropriate integration algorithm to get a very accurate evaluation. For example, we can use Gauss quadrature integral approximation associated with the Laguerre polynomials [5]. In this approximation, we start by defining the following tridiagonal and symmetric quadrature matrix, which is obtained by using the recursion relation and orthogonality of the Laguerre polynomials as

$$J_{n,m}=\langle n|y|m\rangle=\left(2n+\nu+1\right)\delta_{n,m}-\sqrt{n\left(n+\nu\right)}\delta_{n,m+1}-\sqrt{\left(n+1\right)\left(n+\nu+1\right)}\delta_{n,m-1}. \quad (5)$$

Then, $\langle n|f(y)|m\rangle=\left[f(J)\right]_{n,m}$. Numerically, however, this formula is not practical especially if the function *f*(*y*) is not simple and/or if the size of the basis is relatively large. The following alternative evaluation is much more appropriate numerically and produces highly accurate results. Let $\left\{\tau_n\right\}_{n=0}^{N-1}$ be the *N* eigenvalues of the *N*×*N* truncated version of the quadrature matrix *J* and let $\left\{\Lambda_{m,n}\right\}_{m=0}^{N-1}$ be the corresponding normalized eigenvectors. Then, one can show that [5]

$$\langle n|f(y)|m\rangle\cong(\Lambda F\Lambda^{\mathrm{T}})_{n,m}, \quad (6)$$

where $F$ is the diagonal matrix whose elements are $F_{n,n}=f(\tau_n)$. Therefore, the basis overlap matrix becomes $\Omega\cong\Lambda W\Lambda^{\mathrm{T}}$, where $W$ is the diagonal matrix with the elements $W_{n,n}=(\tau_n)^{2\alpha-\nu-a}e^{-2b\tau_n}$. We show below that exact solution for the continuous energy wave function is written in terms of the two-parameter Meixner-Pollaczek polynomial or the three-parameter continuous dual Hahn polynomial [6]. Whereas, the discrete energy bound states are written in terms of the discrete version of these polynomials. These orthogonal polynomials are presented in Appendix D. In the following two subsections, we address the two cases corresponding to (3a) and (3b) separately.

## 2.1 The first Laguerre basis case (3a)

In this case, $2\alpha=1+\nu-a$ and Eq. (A7a) in Appendix A gives the following matrix representation of the wave operator

$$\begin{aligned}\frac{8}{\lambda^2}\mathcal{J}_{n,m}&=\left[\left(2n+\nu+1\right)\left(4u_0+b^2+1\right)+2ab+4u_1\right]\delta_{n,m}\\&-\left(4u_0+b^2-1\right)\left[\sqrt{n\left(n+\nu\right)}\delta_{n,m+1}+\sqrt{\left(n+1\right)\left(n+\nu+1\right)}\delta_{n,m-1}\right]\end{aligned} \quad (7)$$

where $u_i=2U_i/\lambda^2$, $\nu^2=\left(1-a\right)^2+4u_2$ and the basis overlap matrix (4) is $\Omega_{n,m}=\langle n|y^{1-2a}e^{-2by}|m\rangle$. Next, we consider the three configuration spaces that correspond to $b=0$ and $a=\left\{0,\frac{1}{2},1\right\}$ separately. The continuous energy scattering states are written in terms of the Meixner-Pollaczek polynomial whereas the discrete energy bound states are written in terms of either the Meixner or Krawtchouk polynomial depending on whether the energy spectrum is infinite or finite, respectively [6].

### 2.1.1: $(a,b)=\left(\frac{1}{2},0\right)$:

In this case, the basis functions are orthonormal (i.e., the basis overlap matrix is unity since $\Omega_{n,m}=\langle n|m\rangle=\delta_{n,m}$) and we obtain the following coordinate transformation and potential function

$$y(x)=\left(\lambda x/2\right)^2 \text{ and } V(x)=\frac{\lambda^2 V_0}{4}x^2+V_1+\frac{4V_2}{\lambda^2}\frac{1}{x^2}, \tag{8}$$

where $\{U_0,U_1,U_2\}=\{V_0,V_1-E,V_2\}$, $\nu^2=\frac{1}{4}+4u_2$, $2\alpha=\nu+\frac{1}{2}$, $2\beta=1$ and thus we require that $u_2\geq -1/16$. This potential corresponds to the 3D isotropic oscillator with oscillator frequency $\lambda\sqrt{V_0/2}$ and $\nu=\ell+\frac{1}{2}$ whereas *x* stands for the radial coordinate *r*. We can freely choose a potential energy datum at $V_1=0$. Hence, substituting into the matrix wave equation (7) yields the following

$$\frac{-\varepsilon}{u_0-\frac{1}{4}}f_n=-(2n+\nu+1)\frac{u_0+\frac{1}{4}}{u_0-\frac{1}{4}}f_n+\sqrt{n(n+\nu)}\,f_{n-1}+\sqrt{(n+1)(n+\nu+1)}\,f_{n+1}, \tag{9}$$

where $\varepsilon=2E/\lambda^2$. This is a symmetric three-term recursion relation for the expansion coefficients of the wavefunction $\{f_n\}$. We compare it to the one associated with the normalized Meixner-Pollaczek polynomial $P_n^{\mu}(z,\theta)$ shown in Eq. (D2) of Appendix D. Therefore, if we write $f_n(E)=f_0(E)Q_n(E)$, then we conclude that $Q_n(E)=P_n^{\frac{\nu+1}{2}}\left(-\varepsilon/2\sqrt{-u_0},\theta\right)$ where $\cos\theta=\frac{u_0+1/4}{u_0-1/4}$ requiring that $V_0<0$, which is the repulsive inverted oscillator that does not support local states. Thus, physically we should take $V_0>0$ making the argument of the polynomial *z* pure imaginary and $\frac{u_0+1/4}{u_0-1/4}\leq -1$ or $\frac{u_0+1/4}{u_0-1/4}\geq 1$ for $0\leq u_0<1/4$ or $u_0>1/4$, respectively. Thus, $\cos\theta\to\pm\cosh\theta$, which is accomplished by the replacement $\theta\to i\theta$ in the recursion relation (D2) leading also to the replacement $\sin\theta\to\sinh\theta$ and $z\to iz$. Moreover, the oscillating factor $e^{in\theta}$ in the definition of the polynomial (D1) becomes a decaying exponential $e^{-n\theta}$ that becomes vanishingly small for large values of *n*. This implies that the system has no continuum scattering states but only discrete bound states. Nonetheless, this is physically expected along with the fact that the number of bound states are infinite. Moreover, the bound states energy spectrum is obtained from the spectrum formula associated with the Meixner-Pollaczek polynomials that reads $z^2=-(n+\mu)^2$ giving the well-known energy spectrum of the isotropic oscillator,

$$E_n=\lambda\sqrt{V_0/2}\left(2n+\ell+\tfrac{3}{2}\right). \tag{10}$$

The bound states wavefunction is written in terms of the Meixner polynomial as

$$\psi(E_m,r)\sim\sum_{n=0}^{\infty}M_n^{\frac{\nu+1}{2}}\left(m;e^{-2\theta}\right)\phi_n(\lambda r), \tag{11}$$

where $M_n^{\mu}(m;\beta)$ is defined in Appendix D by Eq. (D7) and $\theta=\cosh^{-1}\frac{u_0+1/4}{u_0-1/4}$.

**2.1.2: $(a,b)=(0,0)$:**

In this case, the basis functions are tri-thogonal since $\Omega_{n,m}=\langle n|y|m\rangle = J_{n,m}$ is a tri-diagonal matrix. With these values of *a* and *b*, we obtain the following coordinate transformation and potential function

$$y(x)=\lambda x \text{ and } V(x)=V_0+\frac{V_1/\lambda}{x}+\frac{V_2/\lambda^2}{x^2}, \tag{12}$$

where $\{U_0,U_1,U_2\}=\{V_0-E,V_1,V_2\}$, $\nu^2=1+4u_2$, $2\alpha=\nu+1$, $2\beta=1$ and we require that $u_2\geq -1/4$. This correspond to the 3D Coulomb problem where *x* stands for the radial coordinate, $V_1=\lambda Z$, and $\nu=2\ell+1$. Here, we can also choose a potential energy reference where $V_0=0$. Using these parameter assignments, we could write the matrix wave equation (7) as

$$\frac{-u_1}{\varepsilon+\frac{1}{4}}f_n=-(2n+\nu+1)\frac{\varepsilon-\frac{1}{4}}{\varepsilon+\frac{1}{4}}f_n+\sqrt{n(n+\nu)}f_{n-1}+\sqrt{(n+1)(n+\nu+1)}f_{n+1}. \tag{13}$$

Comparing this with the recursion relation (D2) and writing $f_n = f_0Q_n$, we conclude that $Q_n(E)=P_n^{\ell+1}\left(-Z/\sqrt{2E},\theta\right)$ where $\cos\theta=\frac{\varepsilon-1/4}{\varepsilon+1/4}$ and $\varepsilon>0$. Thus, the continuous energy wavefunction becomes

$$\psi(E,r)= f_0(E)\sum_{n=0}^{\infty}P_n^{\ell+1}\left(-Z/\sqrt{2E},\theta\right)\phi_n(\lambda r), \tag{14}$$

where $f_0(E)$ is obtained by normalization as the square root of the positive weight function given by Eq. (D3). The scattering phase shift, which is obtained from the asymptotics of this polynomial, is given in Appendix D by Eq. (D5) resulting in the following phase shift for this problem

$$\delta(E)=\arg\Gamma(\ell+1-\mathrm{i}Z/k)+\tfrac{\pi}{2}\left(\ell+\tfrac{1}{2}\right), \tag{15}$$

which is the well-known phase shift for the Coulomb problem with $E=\frac{1}{2}k^2$ [7]. The bound states energy spectrum is obtained from the associated spectrum formula $z^2 = -(n+\mu)^2$ giving the well-known energy spectrum of the Coulomb potential,

$$E_n=-Z^2/2(n+\ell+1)^2. \tag{16}$$

The $m^{\text{th}}$ bound state wavefunction $\psi(E_m,r)$ will be written in terms of the Meixner polynomial, which is the discrete version of the Meixner-Pollaczek polynomial, as

$$\psi(E_m,r)\sim\sum_{n=0}^{\infty}M_n^{\ell+1}\left(m;e^{-2\theta}\right)\phi_n(\lambda r). \tag{17}$$

**2.1.3: $(a,b)=(1,0)$:**

In this case, the basis overlap matrix is not tridiagonal but a full since $\Omega_{n,m}=\langle n|y^{-1}|m\rangle = J^{-1}_{n,m}$. Thus, to maintain the tridiagonal structure of $\mathcal{J}$, it must be eliminated from the matrix wave equation by an energy dependent term in the kinetic energy part of the Hamiltonian. As explained above, this means that we must choose the basis parameter $\nu$ such that $\nu^2=8(V_{const}-E)/\lambda^2$. Moreover, we obtain the following coordinate transformation and potential function

$$y(x)=e^{\lambda x} \text{ and } V(x)=V_0e^{2\lambda x}+V_1e^{\lambda x}+V_2, \tag{18}$$

where $-\infty < x < +\infty$, $\{U_0, U_1, U_2\} = \{V_0, V_1, V_2 - E\}$, $\nu^2 = 8(V_2 - E)/\lambda^2$, $2\alpha = \nu$ and $2\beta = 1$. This corresponds to the 1D Morse potential and to force the potential to vanish as $x \to -\infty$ we choose $V_2 = 0$. The basis will be energy dependent via the dependence of the basis parameter $\nu$ on the energy and the solution is valid only for negative energy since $\nu^2 = -4\varepsilon$. Substitution in the matrix wave equation (7) yields the following

$$\frac{u_1}{u_0 - \frac{1}{4}} f_n = -(2n+\nu+1)\frac{u_0 + \frac{1}{4}}{u_0 - \frac{1}{4}} f_n + \sqrt{n(n+\nu)}\, f_{n-1} + \sqrt{(n+1)(n+\nu+1)}\, f_{n+1}, \quad (19)$$

Comparing this with the recursion relation of the normalized Meixner-Pollaczek polynomial $P_n^{\mu}(z,\theta)$ shown in Appendix D and writing $f_n = f_0 Q_n$, we conclude that $Q_n(E) = P_n^{\frac{\nu+1}{2}}\left(u_1/2\sqrt{-u_0}, \theta\right)$ where $\cos\theta = \frac{u_0 + 1/4}{u_0 - 1/4}$ and thus $V_0 < 0$, which is the inverted Morse potential that does not support localized states for negative energy. As explained in subsection 2.1.1 above, this implies that we should take $V_0 > 0$ and for negative energy the system has no continuum scattering states but only discrete bound states and only if we take $V_1 < 0$. This, of course, is physically expected along with the fact that the number of bound states is finite. The spectrum formula of the Meixner-Pollaczek polynomial gives the bound states energies as

$$E_n = -\tfrac{1}{2}\lambda^2\left(n + \tfrac{1}{2} + u_1/2\sqrt{u_0}\right)^2, \quad (20)$$

where $n = 0,1,..,N$ and $N$ is the larger integer less than or equal to $-u_1/2\sqrt{u_0} - \frac{1}{2}$. Consequently, the bound states wavefunction should be written in terms of the Krawtchouk polynomial not in terms of the Meixner polynomial since the latter has an infinite spectrum. Thus, we write

$$\psi(E_m, x) \sim \sum_{n=0}^{N} K_n^N(m;\beta)\,\phi_n(\lambda x), \quad (21)$$

where $K_n^N(m;\beta)$ is defined in Appendix D by Eq. (D9) and $\beta$ is the non-integer part of $-u_1/2\sqrt{u_0} - \frac{1}{2}$. We show in subsection 2.2.1 below how to obtain the most general solution of the 1D Morse potential that involves the continuous as well as the discrete energy spectrum. However, the negative energy bound state solution obtained here is for a general positive potential strength $V_0$ whereas in subsection 2.2.1 it is fixed as $V_0 = \lambda^2/8$.

**2.1.4: non-tridiagonal representation:**

In this subsection, we consider sample problems with non-tridiagonal matrix wave operators. Nonetheless, the Hamiltonian matrix, which is obtained by substituting $E = 0$ in the wave operator matrix (7), is still tridiagonal. That is, the matrix elements $H_{n,m}$ are on the right-hand side of Eq. (7) above with $U_i = V_i$. On the other hand, the basis overlap matrix $\Omega$ in these problems is not tridiagonal and cannot be eliminated by a counter term from the kinetic energy matrix. We present two examples; one of them is exactly realizable but does not belong to the traditional class of exactly solvable problems whereas the other is numerically solvable. The first one corresponds to $b = 0$ and $a = -\frac{1}{2}$ giving

$$y(x)=\left(\tfrac{3}{2}\lambda x\right)^{2/3} \text{ and } V(x)=V_0 y^{-1}+V_1 y^{-2}+V_2 y^{-3}, \tag{22}$$

where $\nu^2=\frac{9}{4}+4u_2$, $2\alpha=\nu+\frac{3}{2}$, $2\beta=1$ and we require that $u_2\ge-\left(3/4\right)^2$. Choosing the basis parameter $\nu=3\left(\ell+\frac{1}{2}\right)$ makes the last term of the potential an orbital term. The overlap matrix becomes $\Omega_{n,m}=\langle n|y^2|m\rangle$ which could be evaluated exactly by applying the recursion relation of the Laguerre polynomials twice giving the following penta-diagonal symmetric matrix

$$\begin{aligned}\Omega_{n,m}&=\left[(2n+\nu+1)^2+(2n+1)(n+\nu+1)-n\right]\delta_{n,m}\\&-2(2n+\nu)\sqrt{n(n+\nu)}\delta_{n,m+1}-2(2n+\nu+2)\sqrt{(n+1)(n+\nu+1)}\delta_{n,m-1}\\&+\sqrt{(n-1)n(n+\nu-1)(n+\nu)}\delta_{n,m+2}+\sqrt{(n+1)(n+2)(n+\nu+1)(n+\nu+2)}\delta_{n,m-2}\end{aligned} \tag{23}$$

In fact, one can show that $\Omega_{n,m}=J^2_{n,m}$, where $J$ is the quadrature matrix (5). The tridiagonal Hamiltonian matrix is obtained from Eq. (7) by taking $E=0$ giving

$$\begin{aligned}\frac{8}{\lambda^2}H_{n,m}&=\left[(2n+\nu+1)(1+4u_0)+4u_1\right]\delta_{n,m}\\&+(1-4u_0)\left[\sqrt{n(n+\nu)}\delta_{n,m+1}+\sqrt{(n+1)(n+\nu+1)}\delta_{n,m-1}\right]\end{aligned} \tag{24}$$

Thus, with all elements of the matrix wave equation $H|f\rangle=E\Omega|f\rangle$ being realized exactly, the problem is labeled as exactly solvable. Table 1 gives the lowest portion of the energy spectrum for a particular set of values of the potential parameters $\{V_0,V_1\}$ and angular momentum $\ell$.

The second example corresponds to $a=0$ and $b=\frac{1}{2}$ giving

$$y(x)=-2\ln\left(1-\tfrac{x}{L}\right), \tag{25}$$

where $0\le x\le L$ and $\lambda=2/L$. Additionally, the potential function becomes

$$V(x)=\left(1-\tfrac{x}{L}\right)^{-2}\left(V_0+V_1y^{-1}+V_2y^{-2}\right), \tag{26}$$

where $\nu^2=1+4u_2$, $2\alpha=\nu+1$, $2\beta=\frac{3}{2}$ and $u_2\ge-1/4$. This logarithmic 1D potential box is known to have an exact solution at zero energy [8]. The basis overlap matrix $\Omega_{n,m}=\langle n|ye^{-y}|m\rangle$ could be evaluated using the Gauss quadrature approximation associated with the Laguerre polynomials as explained above giving $\Omega=\Lambda W\Lambda^{\mathrm{T}}$, where the diagonal matrix $W$ has the elements $W_{n,n}=\tau_n e^{-\tau_n}$. The Hamiltonian matrix is obtained from Eq. (7) as follows

$$\begin{aligned}\frac{8}{\lambda^2}H_{n,m}&=\left[(2n+\nu+1)\left(\tfrac{5}{4}+4u_0\right)+4u_1\right]\delta_{n,m}\\&+\left(\tfrac{3}{4}-4u_0\right)\left[\sqrt{n(n+\nu)}\delta_{n,m+1}+\sqrt{(n+1)(n+\nu+1)}\delta_{n,m-1}\right]\end{aligned} \tag{27}$$

Using the matrices $H$ and $\Omega$ in the eigenvalue matrix equation $H|f\rangle=E\Omega|f\rangle$ we obtain the lowest part of the energy spectrum shown in Table 2 for a given set of physical parameters.

Finally, we note that all potential functions in this class that correspond to $b=0$ and for various values of $a$ such that either $a<1$ or $a=2,4,6,...=2n$ are all power-law potentials of the form $V(x)=V_0 y^{2a}+V_1 y^{2a-1}$ with $y(x)=\left[(1-a)\lambda x\right]^{\frac{1}{1-a}}$. The $V_2$ potential term in this class becomes an orbital term by choosing the basis parameter $\nu^2=4(1-a)^2\left(\ell+\frac{1}{2}\right)^2$. These potentials are known to have exact solutions at zero energy [9]. Nonetheless, by calculating the non-tridiagonal overlap matrix $\Omega_{n,m}=\langle n|y^{1-2a}|m\rangle$ (e.g., using Gauss quadrature) and using the exact tridiagonal Hamiltonian matrix obtained from Eq. (7) one can obtain the energy spectrum (if they exist) to a very high accuracy. On the other hand, all potential functions in this class that corresponds to $a=0$ are 1D logarithmic of the form $V(x)=\left(1-b\lambda x\right)^{-2}\left(V_0+V_1 y^{-1}+V_2 y^{-2}\right)$ where $y(x)=-b^{-1}\ln\left(1-b\lambda x\right)$ and either $x\geq 0$ for negative $b$ or $0\leq x\leq 1/b\lambda$ for positive $b$. These potentials are also known to be exactly solvable at zero energy [8]. The basis overlap matrix in this case is $\Omega_{n,m}=$ $\langle n|y\,e^{-2by}|m\rangle$.

## 2.2 The second Laguerre basis case (3b)

In this case, $2\alpha=2+\nu-a$ and Eq. (A7b) in Appendix A gives the following tridiagonal matrix representation of the wave operator

$$\begin{aligned}\frac{2}{\lambda^2}\mathcal{J}_{n,m}=&\left[\left(2n+\nu+1\right)\left(n+\tfrac{\nu+ab}{2}+1+u_1\right)-n+u_2+\frac{(1-a)^2}{4}-\frac{(\nu+1)^2}{4}\right]\delta_{n,m}\\&-\left(n+\tfrac{\nu+ab}{2}+u_1\right)\sqrt{n\left(n+\nu\right)}\,\delta_{n,m+1}-\left(n+\tfrac{\nu+ab}{2}+1+u_1\right)\sqrt{\left(n+1\right)\left(n+\nu+1\right)}\,\delta_{n,m-1}\end{aligned}\qquad(28)$$

where $u_i=2U_i/\lambda^2$ and $4u_0=1-b^2$. Aside from the requirement that it is greater than $-1$, the basis parameter $\nu$ is arbitrary to be fixed by physical constraints for a given problem. The basis overlap matrix (4) is $\Omega_{n,m}=$ $\langle n|y^{2-2a}e^{-2by}|m\rangle$ and we consider the three configuration spaces that correspond to $b=0$ and $a=\left\{1,\frac{1}{2},0\right\}$ separately. The continuous energy scattering states are written in terms of the continuous dual Hahn polynomial whereas the discrete energy bound states (if they exist) are written in terms of the dual Hahn polynomial for a finite spectrum.

### 2.2.1: $(a,b)=(1,0)$:

In this case, the basis elements are orthonormal (i.e., $\Omega_{n,m}=\langle n|m\rangle=\delta_{n,m}$) and we obtain the following coordinate transformation and potential function

$$y(x)=e^{\lambda x}\ \text{and}\ V(x)=\frac{\lambda^2}{8}e^{2\lambda x}+V_1 e^{\lambda x}+V_2,\qquad(29)$$

where $-\infty<x<+\infty$, $\left\{U_0,U_1,U_2\right\}=\left\{\lambda^2/8,V_1,V_2-E\right\}$, $2\alpha=\nu+1$ and $2\beta=1$. This potential is the 1D Morse problems. Choosing $V_2=0$ and substituting in the matrix wave equation (28), we obtain the following three-term recursion relation for the expansion coefficients of the wavefunction

$$
\begin{aligned}
\varepsilon f_n &= \left[\left(2n+\nu+1\right)\left(n+\tfrac{\nu}{2}+1+u_1\right)-n-\frac{(\nu+1)^2}{4}\right]f_n \\
&-\left(n+\tfrac{\nu}{2}+u_1\right)\sqrt{n\left(n+\nu\right)}\,f_{n-1}-\left(n+\tfrac{\nu}{2}+1+u_1\right)\sqrt{\left(n+1\right)\left(n+\nu+1\right)}\,f_{n+1}
\end{aligned} \tag{30}
$$

where $\varepsilon = 2E/\lambda^2$ and $1 \geq \nu > -1$ to be fixed by the number of bound states $N$ as $\nu = -2\left(N+1+u_1\right)$ where $N$ is the largest integer less than or equal to $-\frac{1}{2}-u_1$. Comparing (30) with the recursion relation of the continuous dual Hahn polynomial $S_n^{\mu}(z^2;\tau,\sigma)$ given by Eq. (D12) in Appendix D, we can make the following parameters assignment

$$
\tau = \sigma = \frac{\nu+1}{2},\ \mu = u_1 + \frac{1}{2},\ z^2 = \varepsilon\,. \tag{31}
$$

Thus, writing $f_n(E) = f_0(E)Q_n(E)$ we obtain $Q_n(E) = S_n^{u_1+\frac{1}{2}}\left(\varepsilon;\frac{\nu+1}{2},\frac{\nu+1}{2}\right)$ with $u_1 \geq -\frac{1}{2}$ for the continuous energy wavefunction. The associated scattering phase shift is obtained from the asymptotics formula of this polynomial obtained in Appendix D by Eq. (D15) giving [10]

$$
\begin{aligned}
\delta(E) &= \arg\left[\Gamma\left(2i\tfrac{k}{\lambda}\right)\Big/\Gamma\left(u_1+\tfrac{1}{2}+i\tfrac{k}{\lambda}\right)\Gamma\left(\tfrac{\nu+1}{2}+i\tfrac{k}{\lambda}\right)^2\right]+\tfrac{\pi}{2}\left(\nu+\tfrac{1}{2}\right) \\
&= \arg\Gamma\left(2i\tfrac{k}{\lambda}\right)-\arg\Gamma\left(u_1+\tfrac{1}{2}+i\tfrac{k}{\lambda}\right)-2\arg\Gamma\left(\tfrac{\nu+1}{2}+i\tfrac{k}{\lambda}\right)+\tfrac{\pi}{2}\left(\nu+\tfrac{1}{2}\right)
\end{aligned} \tag{32}
$$

The bound states energies are obtained from the spectrum formula of the continuous dual Hahn orthogonal polynomial $z^2 = -(n+\mu)^2$ with $n = 0,1,...,N$ where $N$ is the largest integer less than or equal to $-\mu$. Consequently, we obtain the well-known energy spectrum of the 1D Morse problem as [11]

$$
2E_n/\lambda^2 = -\left(n+u_1+\tfrac{1}{2}\right)^2, \tag{33}
$$

where $n \leq N$ and $N \leq -\left(u_1+\frac{1}{2}\right)$ which requires that the potential strength parameter $V_1$ satisfy $u_1 \leq -\frac{1}{2}$. The $m^{\text{th}}$ bound state wavefunction $\psi(E_m,x)$ will be written in terms of the dual Hahn polynomial, which is the discrete version of the continuous dual Hahn polynomial, as

$$
\psi(E_m,x) \sim \sum_{n=0}^{N} R_n^N\left(m;\tau,\sigma\right)\phi_n(\lambda x)\,, \tag{34}
$$

where $R_n^N\left(m;\tau,\sigma\right)$ is defined in Appendix D by Eq. (D17). In contrast to the limited solution found in subsection 2.1.3 above, which is restricted to the discrete bound states, we obtain here the most general solution of the 1D Morse potential for all energies, continuous and discrete. This is because for a certain range of parameters, the continuous dual Hahn polynomial $S_n^{\mu}(z^2;\tau,\sigma)$ has a continuous as well as a discrete spectrum whereas the Krawtchouk polynomial $K_n^N\left(m;\beta\right)$ has only a discrete spectrum. If $\mu < 0$ such that $\tau$ and $\sigma$ are greater than $-\mu$ then $S_n^{\mu}(z^2;\tau,\sigma)$ has a mix of continuous and discrete spectra, which is reflected in its orthogonality relation that is made up of a continuous integral part and a discrete finite sum part, which reads as follows

$$
\begin{aligned}
&\int_0^{\infty}\rho^{\mu}(z;\tau,\sigma)\,S_n^{\mu}(z^2;\tau,\sigma)\,S_{n'}^{\mu}(z^2;\tau,\sigma)\,dz + 2\frac{\Gamma(\tau-\mu)\Gamma(\sigma-\mu)}{\Gamma(\tau+\sigma)\Gamma(1-2\mu)}\times \\
&\sum_{m=0}^{N}\left(-\mu-m\right)\frac{(\mu+\tau)_m(\mu+\sigma)_m(1-2\mu-m)_m}{(\tau-\mu-m)_m(\sigma-\mu-m)_m m!}\,S_n^{\mu}\left(-(m+\mu)^2;\tau,\sigma\right)S_{n'}^{\mu}\left(-(m+\mu)^2;\tau,\sigma\right) = \delta_{n,n'}
\end{aligned} \tag{35}
$$

where $S_n^{\mu}\left(-(m+\mu)^2;\tau,\sigma\right)$ is proportional to the discrete dual Hahn polynomial $R_n^N\left(m;\hat{\tau},\hat{\sigma}\right)$ where the parameters $\left\{N,\hat{\tau},\hat{\sigma}\right\}$ have simple linear relations to $\left\{\mu,\tau,\sigma\right\}$.

**2.2.2: $(a,b)=\left(\frac{1}{2},0\right)$:**

In this case the basis elements are tri-thogonal since $\Omega_{n,m}=\langle n|y|m\rangle=J_{n,m}$ is tridiagonal and we obtain the following physical configuration

$$y(x)=\left(\lambda x/2\right)^2 \text{ and } V(x)=\frac{\lambda^4}{32}x^2+V_1+\frac{4V_2}{\lambda^2}\frac{1}{x^2}, \tag{36}$$

where $\left\{U_0,U_1,U_2\right\}=\left\{\lambda^2/8,V_1-E,V_2\right\}$, $2\alpha=\nu+\frac{3}{2}$ and $2\beta=1$. This problem corresponds to the 3D isotropic oscillator with oscillator frequency $\frac{1}{4}\lambda^2$ and $x\to r$. We also choose $4u_2=\ell(\ell+1)$ and $V_1=0$ making the wave equation equivalent to the following symmetric three-term recursion relation

$$\begin{aligned}-\frac{1}{4}\left(\ell+\tfrac{1}{2}\right)^2 f_n &= \left[\left(2n+\nu+1\right)\left(n+\tfrac{\nu}{2}+1-\varepsilon\right)-n-\frac{(\nu+1)^2}{4}\right]f_n \\ &-\left(n+\tfrac{\nu}{2}-\varepsilon\right)\sqrt{n\left(n+\nu\right)}\,f_{n-1}-\left(n+\tfrac{\nu}{2}+1-\varepsilon\right)\sqrt{\left(n+1\right)\left(n+\nu+1\right)}\,f_{n+1}\end{aligned} \tag{37}$$

Comparing this with Eq. (D12) and writing $f_n(E)=f_0(E)Q_n(E)$, we obtain $Q_n(E)=S_n^{\frac{1}{2}-\varepsilon}\left(z^2;\frac{\nu+1}{2},\frac{\nu+1}{2}\right)$ where $z^2=-\frac{1}{4}\left(\ell+\frac{1}{2}\right)^2$. This implies that $z$ is pure imaginary, which means that there are no continuum scattering states for this system but only discrete bound states. This, of course is expected for the harmonic oscillator problem. The energy spectrum formula $z^2=-(n+\mu)^2$ for $\mu<0$ (i.e., $\varepsilon>\frac{1}{2}$) yields the well-known spectrum of the 3D isotropic oscillator $E_n=\frac{1}{4}\lambda^2\left(2n+\ell+\frac{3}{2}\right)$ but only for $n=0,1,...,N$ where $N\geq\frac{1}{2}\ell$. Therefore, this solution, which is written here in terms of the discrete dual Hahn polynomial, is limited compared to the general solution found in subsection 2.1.1 in terms of the discrete Meixner polynomial with an infinite spectrum.

**2.2.3: $(a,b)=(0,0)$:**

In this case, the basis overlap matrix is not tridiagonal but penta-diagonal since $\Omega_{n,m}=\langle n|y^2|m\rangle$, which is given by Eq. (23). As explained above, to eliminate the non-tridiagonal components we must require $\lambda^2=8(V_0-E)$. Moreover, $2\alpha=2+\nu$, $2\beta=1$ and we obtain the coordinate transformation and potential function given by Eq. (12). That is,

$$y(x)=\lambda x \text{ and } V(x)=V_0+\frac{V_1/\lambda}{x}+\frac{V_2/\lambda^2}{x^2}, \tag{38}$$

where $\left\{U_0,U_1,U_2\right\}=\left\{\lambda^2/8,V_1,V_2\right\}$ and we choose $V_0=0$. Since $\lambda$ is energy dependent ($\lambda^2=-8E$) whereas the potential function is not, then we must make the parameter replacement $V_1\to\lambda Z$ and $V_2\to\frac{1}{2}\lambda^2\ell(\ell+1)$. Thus, the problem turns into the 3D Coulomb problem for only negative energies since $\lambda^2=-8E$. The wave operator (28) gives a matrix wave equation, which is equivalent to the following recursion relation

$$\frac{(\nu+1)^2}{4} f_n = \left[ \left(2n+\nu+1\right)\left(n+\tfrac{\nu}{2}+1+\tfrac{Z}{\sqrt{-2E}}\right) - n + \left(\ell+\tfrac{1}{2}\right)^2 \right] f_n$$
$$-\left(n+\tfrac{\nu}{2}+\tfrac{Z}{\sqrt{-2E}}\right)\sqrt{n\left(n+\nu\right)}\, f_{n-1} - \left(n+\tfrac{\nu}{2}+1+\tfrac{Z}{\sqrt{-2E}}\right)\sqrt{\left(n+1\right)\left(n+\nu+1\right)}\, f_{n+1} \tag{39}$$

which is identified with that of the continuous dual Hahn polynomial. However, this case does not add any new features to the 3D Coulomb problem studied in subsection 2.1.2 above. In fact, the solution space here is restricted to the negative energy bound states and only for a finite spectrum. Thus, we will not pursue the solution space of the recursion relation (39).

**2.2.4: non-tridiagonal representation:**

We present in this subsection a system that does not belong to the conventional class of exactly solvable potentials. However, we can still produce an exact Hamiltonian matrix leading to a very accurate solution. The Hamiltonian matrix of the system is tridiagonal and symmetric but the basis overlap matrix is not. As an illustration, we choose $a=b=\frac{1}{2}$ and using the results in Appendix C we obtain

$$\lambda x = \sqrt{2}\gamma\left(\tfrac{1}{2},\tfrac{1}{2}y\right), \tag{40}$$

where $0 \le x \le L$, $\lambda = \sqrt{2\pi}/L$ and $\gamma(x,y)$ is the lower incomplete gamma function. Moreover, $2\alpha = \nu + \frac{3}{2}$, $2\beta = \frac{3}{2}$ and the potential function becomes

$$2V(x)/\lambda^2 = \left(\tfrac{3}{16}y + u_1 + u_2 y^{-1}\right)e^{y}. \tag{41}$$

In principle, the coordinate transformation $x(y)$ in (40) could be inverted to give $y(x)$. Figure 1 is a plot of $y(x)$ whereas Figure 2 is a graph of this 1D potential box obtained by varying one of the potential parameters while keeping the rest fixed. The overlap matrix becomes $\Omega_{n,m} = \langle n | y e^{-y} | m \rangle$ that could be evaluated using the Gauss quadrature approximation associated with the Laguerre polynomials as explained above giving $\Omega = \Lambda W \Lambda^{\mathrm{T}}$, where the diagonal matrix $W$ has the elements $W_{n,n} = \tau_n e^{-\tau_n}$. The Hamiltonian matrix is obtained from Eq. (28) as follows

$$\frac{2}{\lambda^2} H_{n,m} = \left[ \left(2n+\nu+1\right)\left(n+\tfrac{\nu}{2}+\tfrac{9}{8}+u_1\right) - n + u_2 + \frac{1}{16} - \frac{(\nu+1)^2}{4} \right]\delta_{n,m}$$
$$-\left(n+\tfrac{\nu}{2}+\tfrac{1}{8}+u_1\right)\sqrt{n\left(n+\nu\right)}\,\delta_{n,m+1} - \left(n+\tfrac{\nu}{2}+\tfrac{9}{8}+u_1\right)\sqrt{\left(n+1\right)\left(n+\nu+1\right)}\,\delta_{n,m-1} \tag{42}$$

Using these matrices in the eigenvalue matrix equation $H|f\rangle = E\Omega|f\rangle$ we obtain the lowest portion of the energy spectrum shown in Table 3 for a given set of physical parameters.

# 3. Solution in the Jacobi basis

Let $y(x)$ be a coordinate transformation such that $-1 \le y \le +1$. A complete set of square integrable functions as basis in the new configuration space with the dimensionless coordinate $y$ has the following elements

$$\phi_n(x) = \mathcal{A}_n (1-y)^{\alpha}(1+y)^{\beta} P_n^{(\mu,\nu)}(y), \tag{43}$$

where $P_n^{(\mu,\nu)}(y)$ is the Jacobi polynomial of degree $n$ in $y$ and the normalization constant is chosen as $\mathcal{A}_n = \sqrt{\frac{2n+\mu+\nu+1}{2^{\mu+\nu+1}} \frac{\Gamma(n+1)\Gamma(n+\mu+\nu+1)}{\Gamma(n+\mu+1)\Gamma(n+\nu+1)}}$. The four dimensionless real parameters are such that $\{\alpha,\beta\}$ are positive whereas $\{\mu,\nu\}$ are greater than $-1$. In Appendix B, we show that the matrix representation of the wave operator, $\mathcal{J} = -\frac{1}{2}\frac{d^2}{dx^2} + V(x) - E$, in this basis is tridiagonal and symmetric if and only if $\frac{dy}{dx} = \lambda(1-y)^a(1+y)^b$ and $V(x) - E = (1-y)^{2a-1}(1+y)^{2b-1}\left[\frac{U_+}{1+y} + \frac{U_-}{1-y} + U_0 + U_1 y\right]$, where $\{U_i\}$ is a set of real parameters. The real dimensionless constants $a$ and $b$ are fixed for each configuration space associated with a given problem. Additionally, we must choose one of three alternative scenarios:

(1) $\begin{cases} 2\alpha = 1+\mu-a \\ 2\beta = 1+\nu-b \end{cases}$, and $\quad U_+ = \frac{\lambda^2}{4}\left[\nu^2 - (1-b)^2\right]$, $U_- = \frac{\lambda^2}{4}\left[\mu^2 - (1-a)^2\right]$. (44a)

(2) $\begin{cases} 2\alpha = 2+\mu-a \\ 2\beta = 1+\nu-b \end{cases}$, and $\quad U_+ = \frac{\lambda^2}{4}\left[\nu^2 - (1-b)^2\right]$, $U_1 = 0$. (44b)

(3) $\begin{cases} 2\alpha = 1+\mu-a \\ 2\beta = 2+\nu-b \end{cases}$, and $\quad U_- = \frac{\lambda^2}{4}\left[\mu^2 - (1-a)^2\right]$, $U_1 = 0$. (44c)

The overlap matrix of the basis elements becomes

$$\begin{aligned} \Omega_{n,m} = \langle \phi_n | \phi_m \rangle = \mathcal{A}_n \mathcal{A}_m \int_0^\infty (1-y)^\mu (1+y)^\nu P_n^{(\mu,\nu)}(y) P_m^{(\mu,\nu)}(y) \times \\ \left[(1-y)^{2\alpha-\mu-a}(1+y)^{2\beta-\nu-b}\right] dy \equiv \langle n | (1-y)^{2\alpha-\mu-a}(1+y)^{2\beta-\nu-b} | m \rangle \end{aligned} \tag{45}$$

where we have defined $\langle y | n \rangle = \mathcal{A}_n (1-y)^{\mu/2}(1+y)^{\nu/2} P_n^{(\mu,\nu)}(y)$. Thus, for the three cases in (44) above, the matrix $\Omega$ becomes

(44a): $\quad \Omega_{n,m} = \langle \phi_n | \phi_m \rangle = \langle n | (1-y)^{1-2a}(1+y)^{1-2b} | m \rangle$.

(44b): $\quad \Omega_{n,m} = \langle \phi_n | \phi_m \rangle = \langle n | (1-y)^{2-2a}(1+y)^{1-2b} | m \rangle$.

(44c): $\quad \Omega_{n,m} = \langle \phi_n | \phi_m \rangle = \langle n | (1-y)^{1-2a}(1+y)^{2-2b} | m \rangle$.

Therefore, it is a tridiagonal matrix only if:

(44a): $\quad (a,b) = \left(\frac{1}{2},\frac{1}{2}\right)$ or $\left(0,\frac{1}{2}\right)$ or $\left(\frac{1}{2},0\right)$.

(44b): $\quad (a,b) = \left(1,\frac{1}{2}\right)$ or $\left(\frac{1}{2},\frac{1}{2}\right)$ or $(1,0)$.

(44c): $\quad (a,b) = \left(\frac{1}{2},1\right)$ or $(0,1)$ or $\left(\frac{1}{2},\frac{1}{2}\right)$.

Similar to the Laguerre case, if $\Omega$ is not tridiagonal then a counter term from the kinetic energy part of the Hamiltonian must be chosen to eliminate it. Now, the analysis of these cases are not as simple as in the Laguerre basis and, thus, we limit our search for them to the general physical configuration corresponding to the class (44a) where $U_1 \neq 0$. This analysis is carried out in Appendix F where it is shown that a non-tridiagonal overlap matrix is allowed while the tridiagonal structure of the wave operator matrix is maintained in the following cases

(44a): $\quad (a,b) = (1,0)$ or $(0,1)$ or $\left(1,\frac{1}{2}\right)$ or $\left(\frac{1}{2},1\right)$ or $(1,1)$.

We should mention that the same set of allowed values of the configuration space parameters $a$ and $b$ derived above is obtained by simply imposing energy independence of the potential function on the above equations for $V(x)-E$. Now, the symmetry of the three cases presented in the set of equations (44) shows that the solution of a problem corresponding to the configuration (*b*,*a*) could easily be obtained from that of (*a*,*b*) with $y\to -y$ and a simple parameter map as follows:

(44a) (*b*,*a*) is obtained from (44a) (*a*,*b*) by:

$y\to -y$, $\mu\leftrightarrow\nu$, $\alpha\leftrightarrow\beta$, $U_+\leftrightarrow U_-$ and $U_1\to -U_1$.

(44c) (*b*,*a*) is obtained from (44b) (*a*,*b*) by:

$y\to -y$, $\mu\leftrightarrow\nu$, $\alpha\leftrightarrow\beta$ and $U_+\to U_-$.

Therefore, if for example we treat the case $(a,b)=(1,0)$ then we do not consider the case $(a,b)=(0,1)$, which is easily obtained from the former by this parameter map. Hence, without any loss of generality this symmetry allows us to limit our study to the following eight distinct physical configurations:

(44a): $(a,b)=\left(\frac{1}{2},\frac{1}{2}\right)$, $\left(0,\frac{1}{2}\right)$, $(1,0)$, $\left(1,\frac{1}{2}\right)$ and $(1,1)$.

(44b): $(a,b)=\left(1,\frac{1}{2}\right)$, $\left(\frac{1}{2},\frac{1}{2}\right)$ and $(1,0)$.

Below, we show that for these physical configurations the exact solution wavefunction is written in terms of orthogonal polynomials in the energy. The three-term recursion relations satisfied by these polynomials are obtained from the tridiagonal matrix representation of the wave operator. However, these polynomials are not found in the mathematics literature and their analytic properties (e.g., weight functions, generating functions, asymptotics, zeros, etc.) are yet to be derived. Nonetheless, they have already been encountered in physics [12-16]. In Appendix E, we define these orthogonal polynomials by their three-term recursion relation that enables us to obtain all of them analytically to any desired order albeit not in closed form. The asymptotics ($n\to\infty$) of these polynomials should all have the same general form

$$P_n^{\mu}(E)\approx n^{-\tau}A^{\mu}(E)\times\cos[n^{\xi}\,\theta(E)+\delta^{\mu}(E)], \qquad (46)$$

where $\tau$ and $\xi$ are real positive constants that depend on the particular polynomial. The studies in [17-19] show that $A^{\mu}(E)$ is the scattering amplitude and $\delta^{\mu}(E)$ is the phase shift. Both depend on the energy and the set of physical parameters $\{\mu\}$. Bound states, if they exist, occur at energies $\{E_m\}$ that make the scattering amplitude vanish, $A^{\mu}(E_m)=0$. The size of this energy spectrum is either finite or infinite. Since the analytic properties of these polynomials are yet to derived, then our exact TRA solution of the corresponding problems do not provide closed form expressions for the energy spectrum or phase shift. Due to the prime significance of these energy polynomials to many physical problems, we hope that experts in the field study them and derive their properties. We lack the expertise to tackle such a problem. Nonetheless, using the known exact energy spectrum of some of the corresponding physical problems we obtain the spectrum formula for special cases of these energy polynomials. However, we will not address here the phase shift associated with the asymptotics of these polynomials, which could also be obtained from the exact scattering solutions of those physical problems. It is only for these mathematical benefits that we elect to study the problems associated with the case (44b) despite the fact that they are restricted versions of those associated with (44a) since $U_1=0$. Note, however, that in (44a) the range of values of the physical parameters $U_\pm$ are constrained to $U_+\ge -\left[\lambda(1-b)/2\right]^2$ and $U_-\ge -\left[\lambda(1-a)/2\right]^2$ whereas in (44b) one of these two constraints is absent.

As in the Laguerre case of the previous section, to solve the matrix wave equation (1), we need to evaluate matrix elements of the form $\langle n|f(y)|m\rangle$, where $f(y)$ is some function and $\langle y|n\rangle$ is defined below Eq. (45). This could be done exactly only for special cases (e.g., $f$ is a simple polynomial in $y$). However, in general and as in the Laguerre basis we can use any appropriate integration routine to get a very accurate evaluation. For example, we can use the Gauss quadrature associated with the Jacobi polynomials [5]. In this approximation, we start by defining the following tridiagonal and symmetric quadrature matrix, which is obtained by using the recursion relation and orthogonality of the Jacobi polynomial as

$$K_{n,m}=\langle n|y|m\rangle=C_n\,\delta_{n,m}+D_{n-1}\,\delta_{n,m+1}+D_n\,\delta_{n,m-1} \tag{47}$$

where $C_n=\frac{\nu^2-\mu^2}{(2n+\mu+\nu)(2n+\mu+\nu+2)}$ and $D_n=\frac{2}{2n+\mu+\nu+2}\sqrt{\frac{(n+1)(n+\mu+1)(n+\nu+1)(n+\mu+\nu+1)}{(2n+\mu+\nu+1)(2n+\mu+\nu+3)}}$. If $\{\tau_n\}_{n=0}^{N-1}$ are the $N$ eigenvalues of the $N\times N$ truncated version of this matrix $K$ and $\{\Lambda_{m,n}\}_{m=0}^{N-1}$ are the corresponding normalized eigenvectors. Then, $\langle n|f(y)|m\rangle\cong(\Lambda F\Lambda^{\mathrm{T}})_{n,m}$, where $F$ is the diagonal matrix whose elements are $F_{n,n}=f(\tau_n)$. Therefore, the basis overlap matrix (45) becomes $\Omega\cong\Lambda W\Lambda^{\mathrm{T}}$, where $W$ is the diagonal matrix with the elements

$$W_{n,n}=(1-\tau_n)^{2\alpha-\mu-a}(1+\tau_n)^{2\beta-\nu-b}. \tag{48}$$

In the following two subsections, we address the two case of Eq. (44a) and Eq. (44b) separately.

## 3.1 The first Jacobi basis case (44a)

In this case, $2\alpha=1+\mu-a$, $2\beta=1+\nu-b$, and Eq. (B7a) gives the following tridiagonal matrix representation of the wave operator

$$\frac{2}{\lambda^2}\mathcal{J}_{n,m}=\tfrac{1}{4}\left[(2n+\mu+\nu+1)^2-(a+b-1)^2+4u_0\right]\delta_{m,n}+u_1\langle m|y|n\rangle, \tag{49}$$

where $u_i=2U_i/\lambda^2$, $\nu^2=(1-b)^2+2u_+$ and $\mu^2=(1-a)^2+2u_-$. The basis overlap matrix (45) is $\Omega_{n,m}=\langle n|(1-y)^{1-2a}(1+y)^{1-2b}|m\rangle$ and in the following subsections, we consider the five distinct configuration space scenarios noted above separately.

### 3.1.1: $(a,b)=\left(\frac{1}{2},\frac{1}{2}\right)$:

In this case, the basis elements are orthogonal since $\Omega_{n,m}=\delta_{n,m}$ and we obtain the following coordinate transformation and potential function

$$y(x)=\sin(\pi x/L)\text{ and }V(x)=V_0+\frac{(V_++V_-)-(V_+-V_-)\sin(\pi x/L)}{\cos^2(\pi x/L)}+V_1\sin(\pi x/L), \tag{50}$$

where $-L/2\le x\le+L/2$ and $\lambda=\pi/L$. Moreover, $\mu^2=\frac{1}{4}+\frac{4}{\lambda^2}V_-$ and $\nu^2=\frac{1}{4}+\frac{4}{\lambda^2}V_+$, which require that $V_\pm\ge-(\pi/4L)^2$. The potential configuration indicates that the problem has only bound states and that the energy spectrum is infinite. Without the last term, this is the trigonometric version of the Scarf potential, which is a member of the conventional class of exactly solvable potentials [11]. Nevertheless, we can still obtain an exact solution with $V_1 \neq 0$ in the TRA and as follows. The matrix wave equation (49) becomes equivalent to the following symmetric three-term recursion relation for the expansion coefficients of the wavefunction

$$\varepsilon f_n = \left[\left(n+\tfrac{\mu+\nu+1}{2}\right)^2 + u_0 + u_1 C_n\right] f_n + u_1\left(D_{n-1} f_{n-1} + D_n f_{n+1}\right), \tag{51}$$

where $u_i = 2V_i/\lambda^2$, $\varepsilon = 2E/\lambda^2$ and the coefficients $C_n$ and $D_n$ are defined below Eq. (47). Writing $f_n(E) = f_0(E)Q_n(E)$ and comparing this recursion relation to that of Eq. (E1) in Appendix E, we conclude that $Q_n(E) = H_n^{(\mu,\nu)}\left(\varepsilon - u_0; u_1\right)$. Unfortunately, this orthogonal polynomial and its discrete version are not found in the mathematics literature. Moreover, the physics literature does not give us a spectrum formula for this energy polynomial since the solution of the potential (50) with $V_1 \neq 0$ is not known analytically. However, the matrix elements of the Hamiltonian and $\Omega$ are known exactly, then in principle we should be able to obtain an exact solution in this tridiagonal representation approach. Table 4 gives the energies of the lowest bound states for a given set of potential parameters. Table 5 shows good agreement between the bound state energies obtained here with $V_0 = V_\pm = 0$ and those listed in Table I of Ref. [14]. Of course, it is easy to see from (51) that with $V_1 = 0$ one obtains the well-known energy spectrum of the trigonometric Scarf potential as $\varepsilon_n = \left(n+\frac{\mu+\nu+1}{2}\right)^2 + u_0$, where $n = 0,1,2,...$ Therefore, we conclude that the spectrum formula of the polynomial $H_n^{(\mu,\nu)}\left(z;\sigma\right)$ for the special case where $\sigma = 0$ is

$$z_n = \left(n+\tfrac{\mu+\nu+1}{2}\right)^2. \tag{52}$$

### 3.1.2: $(a,b)=\left(0,\frac{1}{2}\right)$:

In this case we obtain a tri-thogonal basis since $\Omega_{n,m} = \langle n|(1-y)|m\rangle = \delta_{n,m} - K_{n,m}$ and $K$ is the tridiagonal symmetric matrix (Jacobi quadrature matrix) given by Eq. (47). The corresponding coordinate transformation and potential function are

$$y(x) = 2\left(x/L\right)^2 - 1 \text{ and } V(x) = \frac{1/4}{1-(x/L)^2}\left[2V_0 + \frac{V_+}{(x/L)^2} + \frac{V_-}{1-(x/L)^2}\right] - V_1\frac{(x/L)^2 - \frac{1}{2}}{(x/L)^2 - 1}, \tag{53}$$

where $0 \le x \le L$ and $\lambda = 2\sqrt{2}/L$. This potential box was never studied before. Figure 3 is a plot of the potential obtained by varying one parameter while keeping the others fixed. The matrix wave equation (49) gives the following equivalent symmetric three-term recursion relation

$$\varepsilon f_n = \left[\left(n+\tfrac{\mu+\nu+1}{2}\right)^2 + u_0 - \tfrac{1}{16} + (\varepsilon+u_1)C_n\right] f_n + (\varepsilon+u_1)\left(D_{n-1} f_{n-1} + D_n f_{n+1}\right), \tag{54}$$

where $u_i = 2V_i/\lambda^2$, $\mu^2 = 1 + \frac{4}{\lambda^2}V_-$ and $\nu^2 = \frac{1}{4} + \frac{4}{\lambda^2}V_+$ which requires that $V_+ \ge -1/2L^2$ and $V_- \ge -2/L^2$. If we write $f_n(E) = f_0(E)Q_n(E)$ and compare this recursion relation with Eq. (E1), we conclude that $Q_n(E) = H_n^{(\mu,\nu)}\left(\varepsilon - u_0 + \frac{1}{16}; \varepsilon + u_1\right)$. The configuration of the potential (53) indicates that the problem has only bound states and that the energy spectrum is infinite. However, since the spectrum formula for the associated discrete orthogonal polynomial is not yet known, we list in Table 6 accurate evaluation of the lowest energy eigenvalues for a particular set of potential parameters $\{V_0,V_1,V_\pm\}$. This is obtained by solving the matrix eigenvalue wave equation $H|f\rangle=E\Omega|f\rangle$, where the tridiagonal Hamiltonian matrix is obtained by substituting $E=0$ in the wave operator matrix (49) and $\Omega$ is obtained explicitly as given above. The bound states wavefunction $\psi_m(x)$ is written as an infinite series in the basis (43) with the discrete polynomials $h_n^{(\mu,\nu)}(z_m;\sigma)$ (mentioned in Appendix E) as expansion coefficients. The spectrum formula (52) for $H_n^{(\mu,\nu)}(z;0)$ implies that there is a class of exact solutions of this problem for the special case where the energy $\varepsilon=-u_1$ such that $-u_1=\left(n+\frac{\mu+\nu+1}{2}\right)^2+u_0-\frac{1}{16}$ for $n=0,1,2,\ldots$

### 3.1.3: $(a,b)=(1,0)$:

In this case the basis overlap matrix is not tridiagonal but, in fact, a full matrix since $\Omega_{n,m}=\langle n|\frac{1+y}{1-y}|m\rangle=\left(\frac{1+K}{1-K}\right)_{n,m}$ where $K$ is the tridiagonal Jacobi quadrature matrix given by Eq. (47). Therefore, to obtain a tridiagonal matrix for the wave operator this matrix must be eliminated by a balancing term in the kinetic energy part of the Hamiltonian as will be shown below. The coordinate transformation and potential function are

$$y(x)=1-2e^{-\lambda x}\ \text{and}\ V(x)=\frac{1}{e^{\lambda x}-1}\left[V_0+V_1\left(1-2e^{-\lambda x}\right)+\frac{V_+/2}{1-e^{-\lambda x}}\right], \tag{55}$$

where $x\geq 0$ and we have chosen $V_-=0$ to force the potential to vanish at infinity (at $y=+1$). Moreover, $U_+=V_+$ and $U_-=-2E$ giving $\nu^2=1+\frac{4}{\lambda^2}V_+$ and $\mu^2=-8E/\lambda^2=$ $-4\varepsilon$ which requires that $V_+\geq-(\lambda/2)^2$ and limits the solution to the negative energy sector. Additionally, the basis become energy dependent via the dependence of the basis parameter $\mu$ on energy. Without the $V_1$ term, this potential could be rewritten as

$$V(x)=\frac{V_+/8}{\sinh^2(\lambda x/2)}+\frac{V_0/2}{\tanh(\lambda x/2)}-\frac{V_0}{2}, \tag{56}$$

which is the hyperbolic Eckart potential (with $\lambda\to\lambda/2$) that belongs to the conventional class of exactly solvable potentials [11]. However, the tridiagonal representation approach gives an exact solution even if $V_1\neq 0$. Now, the configuration of the potential (55) indicates that the problem has continuum scattering states and possibly a finite number of bound states. The solution obtained here is only for the latter since reality requires that the energy be negative. The matrix wave equation (49) becomes equivalent to the following three-term recursion relation for the expansion coefficients of the wave function

$$-u_0 f_n=\left[\left(n+\frac{\mu+\nu+1}{2}\right)^2+u_1C_n\right]f_n+u_1\left(D_{n-1}f_{n-1}+D_nf_{n+1}\right), \tag{57}$$

where $u_i=2V_i/\lambda^2$. Now, Writing $f_n=f_0Q_n$ and comparing this recursion relation with Eq. (E1), we conclude that $Q_n(E)=H_n^{(\mu,\nu)}(-u_0;u_1)$. Since the spectrum formula for this polynomial with $u_1\neq 0$ is not known yet, then we have to resort to numerical techniques to calculate the energy spectrum of the potential (55) for a given set of parameters. Now, because the energy variable does not appear explicitly in the matrix wave equation (57), then finding the energy spectrum is not as straightforward as calculating the eigenvalues of a matrix. Nonetheless, the tridiagonal Hamiltonian matrix could be obtained from (49) by setting $E=0$ and, consequently, $\mu=0$ giving

$$\frac{2}{\lambda^2}H_{n,m}=\left[\left(n+\tfrac{\nu+1}{2}\right)^2+u_0+u_1C_n\right]\delta_{n,m}+u_1\left(D_{n-1}\delta_{n,m+1}+D_n\delta_{n,m-1}\right), \tag{58}$$

where $u_i=2V_i/\lambda^2$. With this Hamiltonian matrix and Ω being explicitly given above, the matrix eigenvalue wave equation $H|f\rangle=E\Omega|f\rangle$ could easily be solved numerically and to the desired accuracy for the energy spectrum. Table 7 is a list of the energy of all bound states (finite in number) for the given set of values of the potential parameters $\{V_+,V_1,V_0\}$. In [16], we did consider this potential as model for an electron interacting with and extended molecule and resulting in a screened Coulomb potential with non-orbital barrier. The bound states wavefunction $\psi_m(x)$ is written as a finite series in the basis (43) with the discrete polynomials $k_n^{(\mu,\nu)}(z_m;\sigma)$ (declared in Appendix E) as expansion coefficients. It is also worth noting that for $V_1=0$ the spectrum formula (52) for $H_n^{(\mu,\nu)}(z;0)$ gives the following equation $-u_0=\left(n+\frac{\mu+\nu+1}{2}\right)^2$, which results in the energy spectrum formula for the potential (56) as $\varepsilon_n=-\left(n+\frac{\nu+1}{2}\pm\sqrt{-u_0}\right)^2$.

### 3.1.4: $(a,b)=\left(1,\frac{1}{2}\right)$:

In this case the basis overlap matrix is not tridiagonal but a full matrix since $\Omega_{n,m}=\langle n|(1-y)^{-1}|m\rangle$ or $\Omega=(1-K)^{-1}$. Therefore, a balancing term in the kinetic energy part of the Hamiltonian must be used to remove it as will be done below. The coordinate transformation and potential function are

$$y(x)=2\tanh^2(\lambda x)-1 \text{ and } V(x)=\frac{V_+}{\sinh^2(\lambda x)}+2\frac{V_0+V_1[2\tanh^2(\lambda x)-1]}{\cosh^2(\lambda x)}, \tag{59}$$

where $x\geq 0$ and we have replaced $\lambda$ by $\lambda\sqrt{2}$ and took $V_-=0$ so that the potential vanishes at infinity ($y=+1$). Moreover, $U_+=V_+$ and $U_-=-E$ giving $\nu^2=\frac{1}{4}+\frac{2}{\lambda^2}V_+$ and $\mu^2=-2E/\lambda^2=-\varepsilon$, which requires that $V_+\geq-\lambda^2/8$ and limits the solution to negative energies. Without the $V_1$ term, this potential is the hyperbolic Pöschl-Teller potential, which belongs to the conventional class of exactly solvable problems [11]. However, it has never been studied in its general form (59). Nonetheless, the TRA can give an exact solution even if $V_1\neq 0$. Now, the configuration of the potential (59) indicates that the problem has continuum scattering states and possibly bound states with finite energy spectrum. The matrix wave equation (49) becomes equivalent to the following recursion relation for the expansion coefficients of the wavefunction

$$-\frac{u_0}{2}f_n=\left[\left(n+\frac{\mu+\nu+1}{2}\right)^2-\frac{1}{16}+\frac{u_1}{2}C_n\right]f_n+\frac{u_1}{2}\left(D_{n-1}f_{n-1}+D_nf_{n+1}\right), \tag{60}$$

where $u_i=2V_i/\lambda^2$. Note that the basis elements are energy dependent since the parameter $\mu$ depends on the energy. Writing $f_n=f_0Q_n$ and comparing this recursion relation with Eq. (E1), we conclude that $Q_n(E)=H_n^{(\mu,\nu)}\left(\frac{1}{16}-u_0/2;u_1/2\right)$. Now, the spectrum formula for this polynomial with $u_1\neq 0$ is not known yet. Then as we did in the previous subsection, we have to resort to numerical techniques to calculate the energy spectrum of the potential (59) for a given set of parameters. The tridiagonal Hamiltonian matrix could be obtained from (49) by setting $E=0$ and, consequently, $\mu=0$ giving

$$\frac{2}{\lambda^2}H_{n,m}=\left[2\left(n+\tfrac{\nu+1}{2}\right)^2-\tfrac{1}{8}+u_0+u_1C_n\right]\delta_{n,m}+u_1\left(D_{n-1}\delta_{n,m+1}+D_n\delta_{n,m-1}\right), \tag{61}$$

With this Hamiltonian matrix and $\Omega$ being given explicitly above, the matrix eigenvalue wave equation $H|f\rangle=E\Omega|f\rangle$ could easily be solved numerically for the energy spectrum. Table 8 gives the energy of all bound states (finite in number) for the given set of values of the potential parameters $\{V_+,V_1,V_0\}$. The bound states wavefunction $\psi_m(x)$ is written as a finite series in the basis (43) with the discrete polynomials $k_n^{(\mu,\nu)}(z_m;\sigma)$ (noted in Appendix E) as expansion coefficients. The spectrum formula (52) for $H_n^{(\mu,\nu)}(z;0)$ gives the following energy spectrum formula for the potential (59) with $V_1=0$

$$\frac{2E_n}{\lambda}=-\left(2n+\nu+1-\sqrt{\tfrac{1}{4}-2u_0}\right)^2, \tag{62}$$

which is identical to that for the hyperbolic Pöschl-Teller potential.

### 3.1.5: $(a,b)=(1,1)$:

In this case, the coordinate transformation and potential function are

$$y(x)=\tanh(\lambda x)\ \text{ and }\ V(x)=W_+-W_-\tanh(\lambda x)+\frac{V_0+V_1\tanh(\lambda x)}{\cosh^2(\lambda x)}, \tag{63}$$

where $-\infty<x<+\infty$, $W_\pm=V_+\pm V_-$ and $U_\pm=V_\pm-\tfrac{1}{2}E$. Without the $V_1$ term, this is the well-known hyperbolic Rosen-Morse potential [11]. The basis overlap matrix $\Omega_{n,m}=\langle\phi_n|\phi_m\rangle=\langle n|(1-y^2)^{-1}|m\rangle$ is a full matrix since $\Omega=\left(1-K^2\right)^{-1}$. As explained above, writing the parameter constraint $U_\pm=V_\pm-\tfrac{1}{2}E$ will eliminate the non-tridiagonal components from the matrix wave operator. This constraint is explicitly written as

$$\nu^2=\tfrac{4}{\lambda^2}\left(V_+-\tfrac{1}{2}E\right)\ \text{ and }\ \mu^2=\tfrac{4}{\lambda^2}\left(V_--\tfrac{1}{2}E\right). \tag{64}$$

Nonetheless, to make the potential vanish at $x\to\pm\infty$ we must choose $V_\pm=0$ giving $\mu^2=\nu^2=-\varepsilon$, which means that the solution obtained here is only for negative energies. Additional, with $V_\pm=0$ the physical system corresponds to the potential

$$V(x)=\frac{V_0+V_1\tanh(\lambda x)}{\cosh^2(\lambda x)}. \tag{65}$$

With $V_1\neq 0$, this system is not an element of the conventional class of exactly solvable potentials. However, it has been studied in [15] where it was referred to as the "hyperbolic single wave" potential. The configuration of this potential indicates that the problem has continuum scattering states and possibly a finite number of bound states. However, the solution obtained here is only for the negative energy bound states. The resulting three-term recursion relation for the expansion coefficients of the wavefunction is obtained from the matrix wave equation (49) as follows

$$-u_0f_n=\left(n+\tfrac{\mu+\nu}{2}\right)\left(n+\tfrac{\mu+\nu}{2}+1\right)f_n+u_1\left(D_{n-1}f_{n-1}+D_nf_{n+1}\right), \tag{66}$$

where $u_i=2V_i/\lambda^2$, $\varepsilon=2E/\lambda^2$ and $\mu=\pm\nu=\sqrt{-\varepsilon}$. That is,

If $\mu=\nu$: $-u_0 f_n=(n+\mu)(n+\mu+1)f_n+\frac{u_1}{2}\left[\sqrt{\frac{n(n+2\mu)}{(n+\mu)^2-1/4}}f_{n-1}+\sqrt{\frac{(n+1)(n+2\mu+1)}{(n+\mu+1)^2-1/4}}f_{n+1}\right]$ (66a)

If $\mu=-\nu$: $-u_0 f_n=n(n+1)f_n+\frac{u_1}{2}\left[\sqrt{\frac{n^2-\mu^2}{n^2-1/4}}f_{n-1}+\sqrt{\frac{(n+1)^2-\mu^2}{(n+1)^2-1/4}}f_{n+1}\right]$, (66b)

Writing $f_n=f_0Q_n$ and comparing the recursion relation (66) to Eq. (E1) with $\mu^2=\nu^2=-\varepsilon$, we conclude that $Q_n(E)=H_n^{(\mu,\nu)}\left(-u_0+\frac{1}{4};u_1\right)$. In the absence of an exact analytic formula for the spectrum of this polynomial with $u_1\neq 0$, we resort to numerical techniques to calculate the energy spectrum of the potential (65) for a given set of parameters. The tridiagonal Hamiltonian matrix could be obtained from (49) by setting $E=0$ and, consequently, $\mu=\nu=0$ giving

$$\frac{2}{\lambda^2}H_{n,m}=\left[n(n+1)+u_0+u_1C_n\right]\delta_{n,m}+u_1\left(D_{n-1}\delta_{n,m+1}+D_n\delta_{n,m-1}\right), \qquad (67)$$

With this Hamiltonian matrix and $\Omega$ being given explicitly as shown above, the matrix eigenvalue wave equation $H|f\rangle=E\Omega|f\rangle$ could easily be solved numerically for the energy spectrum. Table 9 is a list of all bound states (finite in number) energies for the given set of values of the potential parameters $\{V_1,V_0\}$. The bound states wavefunction $\psi_m(x)$ is written as a finite series in the basis (43) with the discrete polynomials $k_n^{(\mu,\nu)}(z_m;\sigma)$ (declared in Appendix E) as expansion coefficients. Moreover, the spectrum formula (52) for $H_n^{(\mu,\nu)}(z;0)$ gives the following energy spectrum for the potential (65) with $V_1=0$

$$\varepsilon_n=-\left(n+\frac{1}{2}-\sqrt{\frac{1}{4}-u_0}\right)^2, \qquad (68)$$

for $V_0<0$ and $n=0,1,..,N$ where $N\leq\sqrt{\frac{1}{4}-u_0}-\frac{1}{2}$. This is identical to the energy spectrum for the special cases of either the hyperbolic Rosen-Morse, Pöschl-Teller or Scarf potentials [11].

**3.1.6: non-tridiagonal representation:**

In this subsection, we give two examples where the matrix wave operator is not tridiagonal but the Hamiltonian matrix is. That is, no kinetic energy counter term could be found to eliminate the non-tridiagonal matrix $\Omega$. The Hamiltonian matrix elements are obtained by substituting $E=0$ in the wave operator matrix, which is the right side of Eq. (49) above with $U_i=V_i$.

The first example corresponds to $(a,b)=\left(\frac{3}{2},\frac{1}{2}\right)$ resulting in the following coordinate transformation and potential function

$$y(x)=\frac{(\lambda x)^2-1}{(\lambda x)^2+1} \text{ and } V(x)=\frac{2}{(\lambda x)^2+1}\left\{V_-+\frac{V_+}{(\lambda x)^2}+\frac{2}{(\lambda x)^2+1}\left[V_0+V_1\frac{(\lambda x)^2-1}{(\lambda x)^2+1}\right]\right\} \qquad (69)$$

where $x\geq 0$, $2\alpha=\mu-\frac{1}{2}$, $2\beta=\nu+\frac{1}{2}$, $\mu^2=\frac{1}{4}+\frac{4}{\lambda^2}V_-$ and $\nu^2=\frac{1}{4}+\frac{4}{\lambda^2}V_+$ requiring that $V_\pm\geq-\left(\lambda/4\right)^2$. Figure 4 is a plot of this potential obtained by varying one of the potential parameters while keeping the rest fixed. The structure of the potential suggests the existence of scattering states and possibly (with the right set of values of the potential parameters) bound states. The basis overlap matrix is $\Omega_{n,m} = \langle n | (1-y)^{-2} | m \rangle = \left[ (1-K)^{-2} \right]_{n,m}$ where $K$ is the Jacobi quadrature matrix (47). The tridiagonal Hamiltonian matrix is obtained from (49) as

$$\frac{2}{\lambda^2} H_{n,m} = \left[ \left( n + \tfrac{\mu+\nu+1}{2} \right)^2 + u_0 - \frac{1}{4} + u_1 C_n \right] \delta_{n,m} + u_1 \left( D_{n-1} \delta_{n,m+1} + D_n \delta_{n,m-1} \right), \qquad (70)$$

where $u_i = 2V_i/\lambda^2$. With this Hamiltonian matrix and $\Omega$ being derived explicitly, the matrix eigenvalue wave equation $H|f\rangle = E\Omega|f\rangle$ could easily be solved numerically and to the desired accuracy for the energy spectrum. Table 10 gives the energy of all finite number of bound states for the given set of values of the potential parameters.

For the second example, we take $(a,b) = \left(-\frac{1}{2}, -\frac{1}{2}\right)$ resulting in the following coordinate transformation and potential function

$$2\lambda x = y\sqrt{1-y^2} + \sin^{-1}(y) \text{ and } V(x) = \left(1-y^2\right)^{-2} \left[ \frac{V_+}{1+y} + \frac{V_-}{1-y} + V_0 + V_1 y \right], \qquad (71)$$

where $-L/2 \le x \le +L/2$, $\lambda = \pi/2L$, $2\alpha = \mu + \frac{3}{2}$, and $2\beta = \nu + \frac{3}{2}$. Moreover, $\mu^2 = \frac{9}{4} + \frac{4}{\lambda^2} V_-$ and $\nu^2 = \frac{9}{4} + \frac{4}{\lambda^2} V_+$ which requires that $V_\pm \ge -\left(3\pi/8L\right)^2$. In principle, the coordinate transformation $x(y)$ in (71) could be inverted to give $y(x)$. Figure 5 is a plot of $y(x)$ whereas Figure 6 is a graph of the potential box obtained by varying one of the potential parameters while keeping the rest fixed. It is clear that the system consists of an infinite number of bound states. The basis overlap matrix is $\Omega_{n,m} = \langle n | \left(1-y^2\right)^2 | m \rangle = \left[ (1-K^2)^2 \right]_{n,m}$ where $K$ is the quadrature matrix (47). The tridiagonal Hamiltonian matrix is obtained from (49) as

$$\frac{2}{\lambda^2} H_{n,m} = \left[ \left( n + \tfrac{\mu+\nu+1}{2} \right)^2 + u_0 - 1 + u_1 C_n \right] \delta_{n,m} + u_1 \left( D_{n-1} \delta_{n,m+1} + D_n \delta_{n,m-1} \right). \qquad (72)$$

Table 11 gives the energy of the lowest bound states for a given set of values of the potential parameters.

## 3.2 The second Jacobi basis case (44b)

In this case, $2\alpha = 2 + \mu - a$, $2\beta = 1 + \nu - b$ and Eq. (B7b) gives the following tridiagonal matrix representation of the wave operator

$$\frac{2}{\lambda^2} \mathcal{J}_{n,m} = \left[ -\frac{2n(n+\nu)}{2n+\mu+\nu} - \frac{\mu+a}{2} \left( \mu - a + 2 \right) + u_- \right] \delta_{m,n}$$
$$+ \left( 2n + \mu + \nu + 1 \right) D_{n-1} \delta_{m,n-1} + \left[ \left( n + \tfrac{\mu+\nu}{2} + 1 \right)^2 - \tfrac{1}{4} \left( a + b - 1 \right)^2 + u_0 \right] \left( \delta_{m,n} - \langle m | y | n \rangle \right) \qquad (73)$$

where $u_i = 2U_i/\lambda^2$ and $\nu^2 = (1-b)^2 + 2u_+$. Aside from the requirement that is greater than $-1$, the basis parameter $\mu$ is arbitrary to be fixed by physical constraints. The basis overlap matrix (45) is $\Omega_{n,m} = \langle n | (1-y)^{2-2a} (1+y)^{1-2b} | m \rangle$ and we consider the three distinct configuration spaces scenarios identified above in the following subsection.

### 3.2.1: $(a,b)=\left(1,\frac{1}{2}\right)$:

In this case, the basis elements are orthonormal since $\Omega_{n,m}=\langle n|m\rangle=\delta_{n,m}$. The coordinate transformation and potential function become

$$y(x)=2\tanh^2(\lambda x)-1 \text{ and } V(x)=\frac{V_+}{\sinh^2(\lambda x)}+\frac{2V_0}{\cosh^2(\lambda x)}, \qquad (74)$$

where $x\geq 0$ and we replaced $\lambda$ by $\lambda\sqrt{2}$. Moreover, $\nu^2=\frac{1}{4}+\frac{2}{\lambda^2}V_+$ while the basis parameter $\mu$ is so far arbitrary. To make the potential vanish at infinity, we took $V_-=0$ and reality of the basis parameter $\nu$ dictates that $V_+\geq-\lambda^2/8$. This potential is the well-known hyperbolic Pöschl-Teller potential and it is a special case of that in subsection 3.1.4 above given by Eq. (59) but with $V_1=0$. However, the solution here is for all energies (continuum scattering states as well as a finite number of bound states) while in subsection 3.1.4 it was restricted only to the negative energy bound states. Thus, we obtain here the orthogonal energy polynomial for the continuum scattering states and use the well-known energy spectrum of the hyperbolic Pöschl-Teller potential to derive its discrete spectrum formula. For that, we write the three-term recursion relation of the coefficients of the continuum wavefunction using the matrix wave operator (73), which gives

$$\begin{aligned}\frac{\varepsilon}{2}f_n=&\left\{-\frac{2n(n+\nu)}{2n+\mu+\nu}-\frac{(\mu+1)^2}{2}+\left[\left(n+\frac{\mu+\nu}{2}+1\right)^2-\frac{1}{16}+\frac{u_0}{2}\right](1-C_n)\right\}f_n\\&-\left[\left(n+\frac{\mu+\nu}{2}\right)^2-\frac{1}{16}+\frac{u_0}{2}\right]D_{n-1}\,f_{n-1}-\left[\left(n+\frac{\mu+\nu}{2}+1\right)^2-\frac{1}{16}+\frac{u_0}{2}\right]D_n\,f_{n+1}\end{aligned} \qquad (75)$$

where $\varepsilon=2E/\lambda^2$, $u_0=2V_0/\lambda^2$ and $\nu^2=\frac{1}{4}+u_+$. Writing $f_n$ as $f_0\,Q_n$ and comparing this recursion relation with Eq. (E3) in Appendix E, we conclude that $Q_n(E)=G_n^{(\mu,\nu)}\left(-\frac{\varepsilon}{2};\frac{u_0}{2}-\frac{1}{16}\right)$. These are the expansion coefficients of the wavefunction in the basis (43) that has only continuous energy states or a combination of continuous and discrete states. The alternative depends on the range of values of the parameters $\{u_0,\mu,\nu\}$. On the other hand, pure bound states will be written in terms of the discrete version of this polynomial, $g_n^{(\mu,\nu)}(z;\sigma)$, which is obtained as follows. First, the energy spectrum is identical to that of the hyperbolic Pöschl-Teller potential, which is well known and reads as follows [11]

$$\frac{2E_n}{\lambda^2}=-\left(2n+\nu+1-\sqrt{\tfrac{1}{4}-2u_0}\right)^2, \qquad (76)$$

where $n=0,1,..,N$ and $N$ is the largest integer less than or equal to $\frac{1}{2}\sqrt{\frac{1}{4}-2u_0}-\frac{\nu+1}{2}$. This physical result implies that the basis parameter $\mu$ has a simple linear dependence on the remaining two physical parameters $\{u_0,\nu,N\}$. Therefore, we conclude that the spectrum formula for the orthogonal polynomial $G_n^{(\mu,\nu)}(z;\sigma)$ is

$$z_n=2\left(n+\frac{\nu+1}{2}-\sqrt{-\sigma}\right)^2, \qquad (77)$$

where $n=0,1,..,N$ and $N$ is the largest integer less than or equal to $\sqrt{-\sigma}-\frac{\nu+1}{2}$ and $\sigma<0$. We could have also obtained the phase shift in the asymptotics of the polynomial $G_n^{(\mu,\nu)}\left(-\frac{\varepsilon}{2};\frac{u_0}{2}-\frac{1}{16}\right)$ had we also used the known exact scattering phase shift associated with the hyperbolic Pöschl-Teller potential in (74).

### 3.2.2: $(a,b)=\left(\frac{1}{2},\frac{1}{2}\right)$:

In this case, the basis is tri-thogonal basis since $\Omega_{n,m}=\langle n|(1-y)|m\rangle=\delta_{n,m}-K_{n,m}$ is a tridiagonal matrix. The coordinate transformation and potential function are

$$y(x)=\sin\left(\pi x/L\right) \text{ and } V(x)=V_0+\frac{(V_++V_-)-(V_+-V_-)\sin(\pi x/L)}{\cos^2(\pi x/L)}, \tag{78}$$

where $-L/2\le x\le +L/2$, $\lambda=\pi/L$ and $\nu^2=\frac{1}{4}+\frac{4}{\lambda^2}V_+$ which requires that $V_+\ge-\left(\pi/4L\right)^2$. The basis parameter $\mu$ is so far arbitrary. This potential is the trigonometric Scarf potential, which is a special case of that given by Eq. (50) in subsection 3.1.1 above with $V_1=0$. However, here the potential parameter $V_-$ is not restricted to be larger than $-\left(\pi/4L\right)^2$. Similarly, here we will not pursue the physical solution of the problem since it is well-known but we use the physical results to obtain the infinite spectrum formula for the associated discrete orthogonal polynomial. The wave operator matrix (73) leads to the following three-term recursion relation

$$\begin{aligned} u_-f_n=&\left\{\frac{2n(n+\nu)}{2n+\mu+\nu}+\frac{(\mu+1)^2-1/4}{2}+\left[\left(n+\frac{\mu+\nu}{2}+1\right)^2+u_0-\varepsilon\right]\left(-1+C_n\right)\right\}f_n \\ &+\left[\left(n+\frac{\mu+\nu}{2}\right)^2+u_0-\varepsilon\right]D_{n-1}f_{n-1}+\left[\left(n+\frac{\mu+\nu}{2}+1\right)^2+u_0-\varepsilon\right]D_nf_{n+1}\end{aligned} \tag{79}$$

where $\nu^2=\frac{1}{4}+2u_+$ and $u_i=2V_i/\lambda^2$. Comparing this with the recursion relation of the orthogonal polynomial $G_n^{(\mu,\nu)}(z;\sigma)$ in Appendix E and writing $f_n/f_0=Q_n$, we conclude that $Q_n(E)=G_n^{(\mu,\nu)}\left(u_-+\frac{1}{8};u_0-\varepsilon\right)$. The spectrum formula of this polynomial given by Eq. (77) produces the following energy spectrum

$$\varepsilon_n=\frac{2E_n}{\lambda^2}=\left(n+\frac{\nu+1}{2}\pm\frac{1}{2}\sqrt{2u_-+\frac{1}{4}}\right)^2+u_0, \tag{80}$$

which is the energy spectrum for the trigonometric Scarf potential (78). Moreover, since $\varepsilon_n\ge u_0$ then the polynomial parameter $\sigma=u_0-\varepsilon$ is negative for all $n$. Thus, $n=0,1,2,..$ and the spectrum is infinite as expected.

### 3.2.3: $(a,b)=\left(1,0\right)$:

In this case also, the basis is tri-thogonal basis since $\Omega_{n,m}=\langle n|(1+y)|m\rangle=\delta_{n,m}+K_{n,m}$ is a tridiagonal matrix. The coordinate transformation and potential function are

$$y(x)=1-2e^{-\lambda x} \text{ and } V(x)=\frac{1}{e^{\lambda x}-1}\left[V_0+\frac{V_+/2}{1-e^{-\lambda x}}\right], \tag{81}$$

where $x\ge0$ and $\nu^2=1+\frac{4}{\lambda^2}V_+$ which requires that $V_+\ge-\left(\lambda/2\right)^2$ and to make the potential vanish at infinity we have chosen $V_-=0$. The basis parameter $\mu$ is so far arbitrary. Note that this potential is a special case of that given by Eq. (55) in subsection 3.1.3 above with $V_1 = 0$ which is the hyperbolic Eckart potential shown in Eq. (56). However, the solution obtained here is for the continuum energy scattering states as well as the finite number of discrete bound states whereas in subsection 3.1.3 the solution was only for the negative energy bound states. The wave operator matrix (73) leads to the following three-term recursion relation

$$\begin{aligned} 2\varepsilon f_n = &\left\{ -\frac{2n(n+\nu)}{2n+\mu+\nu} - \frac{(\mu+1)^2}{2} + \left[ \left( n + \tfrac{\mu+\nu}{2} + 1 \right)^2 + \varepsilon + u_0 \right] (1 - C_n) \right\} f_n \\ &- \left[ \left( n + \tfrac{\mu+\nu}{2} \right)^2 + \varepsilon + u_0 \right] D_{n-1} f_{n-1} - \left[ \left( n + \tfrac{\mu+\nu}{2} + 1 \right)^2 + \varepsilon + u_0 \right] D_n f_{n+1} \end{aligned} \tag{82}$$

where $u_i = 2V_i/\lambda^2$. Comparing this with the recursion relation of the orthogonal polynomial $G_n^{(\mu,\nu)}(z;\sigma)$ in Appendix E while writing $f_n/f_0 = Q_n$, we conclude that $Q_n(E) = G_n^{(\mu,\nu)}(-2\varepsilon; \varepsilon + u_0)$. Now, the spectrum formula (77) for this polynomial gives $\varepsilon_n = -\left( n + \frac{\nu+1}{2} - \sqrt{-\varepsilon_n - u_0} \right)^2$ leading to the following energy spectrum

$$\varepsilon_n = -\frac{1}{4} \left( n + \frac{\nu+1}{2} + \frac{u_0}{n + \frac{\nu+1}{2}} \right)^2, \tag{83}$$

which agrees with the energy spectrum of the hyperbolic Eckart potential of Eq. (81) or Eq. (56). Here again, we could have also obtained the phase shift in the asymptotics of the polynomial $G_n^{(\mu,\nu)}(-2\varepsilon; \varepsilon + u_0)$ had we also used the known exact scattering phase shift associated with the hyperbolic Eckart potential in (81).

Finally, we end this section by making the following important comment. There are additional potential functions in the Jacobi basis class that are exactly solvable in the TRA. This is because we did not exhaust all possible coordinate transformations for a given pair $(a,b)$. For example, in the case $(a,b) = \left(\frac{1}{2}, \frac{1}{2}\right)$ of subsection 3.1.1 we could have also used $y(x) = 2\sin^2(\pi x/2L) - 1$ with $0 \le x \le L$ and $\lambda = \pi/2L$ leading to the following potential function

$$V(x) = V_0 + \frac{V_+/2}{\sin^2(\pi x/2L)} + \frac{V_-/2}{\cos^2(\pi x/2L)} - V_1 \cos(\pi x/L). \tag{84}$$

This potential is a generalization of the trigonometric Pöschl-Teller potential by the additional $V_1$ term (note the doubling of the angle in the new term). We expect that the total number of solvable potentials in this class is double what is shown in Table 12.

# 4. Conclusion

In this work, we have shown that the algebraic Tridiagonal Representation Approach (TRA) gives exact solutions for more problems than those obtained by traditional methods. Table 12 is a list of some of these potentials. They are generalized versions of known potentials except for the new potential box listed in the second row of the Table. Generalization is accomplished by adding a new term (the $V_1$ term). Moreover, the two potentials in the second and third row of Table 12 were never studied in the published literature. Additionally, we have demonstrated that the TRA could also be used to give highly accurate solutions for other problems in which the matrix representation of the wave operator is not tridiagonal but its matrix elements are obtained exactly. The first step in the TRA is to expand the wavefunction in a complete set of square integrable basis elements that satisfy the boundary conditions of configuration space. Thereafter, we choose potential functions and basis parameters such that the matrix representation of the wave operator in this basis is tridiagonal and symmetric. As a result, the matrix wave equation becomes a symmetric three-term recursion relation for the expansion coefficients of the wavefunction, which is then solved exactly in terms of orthogonal polynomials. The physical properties of the system is then extracted in a simple and straightforward manner from the properties of these polynomials. For example, the asymptotics of these polynomials give the bound states energy spectrum and the scattering states phase shift. Another advantage of the TRA is that the representation of the wavefunction as a single bounded sum contains all the energy spectrum of the associated problem both continuous and discrete. This was shown in cases where the energy polynomial has both spectra in its orthogonality relation, which is made up of an integration part over a continuous range of energy and a summation part over a discrete set of energies. Such is the case for the contiguous dual Hahn polynomial (e.g., in the 1D Morse potential of subsection 2.2.1) and the energy polynomial $G_n^{(\mu,\nu)}(z;\sigma)$ (e.g., for the problems of sub-sections 3.2.1 and 3.2.3).

In the process of tackling problems in the Jacobi basis of section 3, the TRA introduced two new orthogonal polynomials that were not treated in the mathematics literature. These were defined in Appendix E by their three-term recursion relations and, for special cases, their spectrum formula were derived. We designated these as $H_n^{(\mu,\nu)}(z;\sigma)$ and $G_n^{(\mu,\nu)}(z;\sigma)$ along with their discrete versions $h_n^{(\mu,\nu)}(z;\sigma)$, $k_n^{(\mu,\nu)}(z;\sigma)$ and $g_n^{(\mu,\nu)}(z;\sigma)$. Due to the significance of these polynomials to many interesting problems in physics, we are hopeful that they will be studies by experts in the field to derive their analytic properties such as the weight function, generating function, orthogonality, differential property, spectrum formula, asymptotics, etc. We do not have the needed skills or expertise to conduct such study.

## Acknowledgements

The Author would like to acknowledge the support by the Saudi Center for Theoretical Physics (SCTP) during the progress of this work.

# Appendix A
# Tridiagonal wave operator matrix in the Laguerre basis

In the transformed configuration space with the dimensionless coordinate $y(x) \geq 0$, a complete set of square integrable basis functions has the elements given by Eq. (2). With a suitable length scale parameter $\lambda$, we can write $\frac{d}{dx} = \lambda\xi(y)\frac{d}{dy}$, where $\xi(y) = \lambda^{-1}\frac{dy}{dx}$. Using the differentiation chain rule, we can write

$$\begin{aligned}
-\frac{2}{\lambda^2}(H-E)\phi_n &= \xi^2\left[\frac{d^2}{dy^2} + \frac{\xi'}{\xi}\frac{d}{dy} - \frac{2}{\lambda^2\xi^2}(V-E)\right]\phi_n \\
&= \mathcal{A}_n y^\alpha e^{-\beta y}\xi^2\left[\left(\frac{d}{dy} + \frac{\alpha}{y} - \beta\right)^2 + \frac{\xi'}{\xi}\left(\frac{d}{dy} + \frac{\alpha}{y} - \beta\right) - \frac{2}{\lambda^2\xi^2}(V-E)\right]L_n^\nu(y) \\
&= \mathcal{A}_n y^\alpha e^{-\beta y}\frac{\xi^2}{y}\left[y\frac{d^2}{dy^2} + \left(2\alpha - 2\beta y + y\frac{\xi'}{\xi}\right)\frac{d}{dy} + \frac{\alpha(\alpha-1)}{y} + \beta^2 y - 2\alpha\beta\right. \\
&\qquad\left. + \frac{\xi'}{\xi}(\alpha - \beta y) - \frac{2y}{\lambda^2\xi^2}(V-E)\right]L_n^\nu(y)
\end{aligned} \tag{A1}$$

where the prime stands for the derivative with respect to $y$. Using the differential equation of the Laguerre polynomial, $\left[y\frac{d^2}{dy^2} + (\nu+1-y)\frac{d}{dy} + n\right]L_n^\nu(y) = 0$, we can write this as

$$\begin{aligned}
-\frac{2}{\lambda^2}(H-E)\phi_n &= \mathcal{A}_n y^\alpha e^{-\beta y}\frac{\xi^2}{y}\left\{\left[2\alpha - \nu - 1 + y\left(1 - 2\beta + \frac{\xi'}{\xi}\right)\right]\frac{d}{dy}\right. \\
&\quad\left. + \frac{\alpha(\alpha-1)}{y} + \beta^2 y - (n + 2\alpha\beta) + \frac{\xi'}{\xi}(\alpha - \beta y) - \frac{2y}{\lambda^2\xi^2}(V-E)\right\}L_n^\nu(y)
\end{aligned} \tag{A2}$$

The tridiagonal requirement and the differential relation of the Laguerre polynomial, $y\frac{d}{dy}L_n^\nu(y) = nL_n^\nu(y) - (n+\nu)L_{n-1}^\nu(y)$, dictate that $y\frac{\xi'}{\xi}$ be linear in $y$. That is, $y\frac{\xi'}{\xi} = a + by$ giving $\xi(y) = y^a e^{by}$, where $a$ and $b$ are real constants. Upon substitution in (A2), we obtain

$$\begin{aligned}
-\frac{2}{\lambda^2}(H-E)\phi_n &= \mathcal{A}_n y^\alpha e^{-\beta y}\frac{\xi^2}{y}\left\{\left[2\alpha + a - \nu - 1 + y(1 + b - 2\beta)\right]\frac{d}{dy}\right. \\
&\quad\left. + \frac{\alpha(\alpha+a-1)}{y} + \beta(\beta-b)y - \left[n + \alpha(2\beta - b) + \beta a\right] - \frac{2y}{\lambda^2\xi^2}(V-E)\right\}L_n^\nu(y)
\end{aligned} \tag{A3}$$

Employing the differential relation of the Laguerre polynomial then multiplying from left by $\phi_m$ and integrating over $x$ while noting that $\lambda\int ...dx = \int ...\xi^{-1}dy$, we obtain the following matrix elements of the wave operator in the basis (2)

$$-\frac{2}{\lambda^2}\langle\phi_m|(H-E)|\phi_n\rangle = \mathcal{A}_n\mathcal{A}_m\int_0^\infty y^\nu e^{-y}\left[y^{2\alpha+a-\nu-2}e^{(1+b-2\beta)y}\right]\times$$

$$L_m^\nu(y)\Big\{\left[2\alpha+a-\nu-1+y(1+b-2\beta)\right]\left[nL_n^\nu(y)-(n+\nu)L_{n-1}^\nu(y)\right] \tag{A4}$$

$$+\left[\alpha(\alpha+a-1)+\beta(\beta-b)y^2-(n+2\alpha\beta+\beta a-\alpha b)y-\frac{2y^2}{\lambda^2\xi^2}(V-E)\right]L_n^\nu(y)\Big\}dy$$

The orthogonality of the Laguerre polynomials, $\mathcal{A}_n^2\int_0^\infty y^\nu e^{-y}L_n^\nu(y)L_m^\nu(y)dy=\delta_{n,m}$, shows that a tridiagonal matrix representation is obtained if and only if $2\beta=1+b$ which dictates that $b>-1$. Writing $2\alpha+a-\nu-2=k$ where $k$ is an integer, we can rewrite (A4) as follows

$$\frac{2}{\lambda^2}\langle\phi_m|(H-E)|\phi_n\rangle=(k+1)\sqrt{n(n+\nu)}\langle m|y^k|n-1\rangle$$

$$-\left[n(k+1)+\tfrac{1}{4}(\nu+k+1)^2-\tfrac{1}{4}(1-a)^2\right]\langle m|y^k|n\rangle \tag{A5}$$

$$+\tfrac{1}{2}(2n+\nu+k+2+ab)\langle m|y^{k+1}|n\rangle-\tfrac{1}{4}(1-b^2)\langle m|y^{k+2}|n\rangle+\frac{2}{\lambda^2}\langle m|y^{k+2}(V-E)/\xi^2|n\rangle$$

where we have defined $\langle y|n\rangle=\mathcal{A}_n y^{\nu/2}e^{-y/2}L_n^\nu(y)$. Therefore, the tridiagonal requirement of the wave operator matrix l two possibilities:

(1) $k=-1$ and $\dfrac{2y}{\lambda^2\xi^2}(V-E)=\dfrac{\nu^2-(1-a)^2}{4y}+A+By$, or (A6a)

(2) $k=0$ and $\dfrac{2y^2}{\lambda^2\xi^2}(V-E)=\tfrac{1}{4}(1-b^2)y^2+A+By$, (A6b)

where $A$ and $B$ are arbitrary dimensionless parameters. Substituting these in (A5) and using the orthogonality of the Laguerre polynomials, we obtain

$$\frac{2}{\lambda^2}\langle\phi_m|(H-E)|\phi_n\rangle=\tfrac{1}{4}(B+b^2-1)\langle m|y|n\rangle+\tfrac{1}{2}(2n+\nu+1+ab+2A)\delta_{m,n}. \tag{A7a}$$

$$\frac{2}{\lambda^2}\langle\phi_m|(H-E)|\phi_n\rangle=\left(n+\tfrac{\nu+ab}{2}+1+B\right)\langle m|y|n\rangle+\sqrt{n(n+\nu)}\delta_{m,n-1}$$

$$-\left[n-A+\frac{(\nu+1)^2}{4}-\frac{(1-a)^2}{4}\right]\delta_{m,n} \tag{A7b}$$

where $\langle n|y|m\rangle$ is the tridiagonal symmetric matrix $J$ given by Eq. (5) (the Laguerre quadrature matrix).

# Appendix B
# Tridiagonal wave operator matrix in the Jacobi basis

In the transformed configuration space with the dimensionless coordinate $-1\le y(x)\le+1$ , a proper and complete set of square integrable basis functions are those given by Eq. (43). Similar to what we have done in Appendix A, we can write the action of the wave operator on this basis functions as follows

$$
\begin{aligned}
-\frac{2}{\lambda^2}(H-E)\phi_n &= \xi^2\left[\frac{d^2}{dy^2}+\frac{\xi'}{\xi}\frac{d}{dy}-\frac{2}{\lambda^2\xi^2}(V-E)\right]\phi_n \\
&= \mathcal{A}_n(1-y)^\alpha(1+y)^\beta\xi^2\left[\left(\frac{d}{dy}-\frac{\alpha}{1-y}+\frac{\beta}{1+y}\right)^2+\frac{\xi'}{\xi}\left(\frac{d}{dy}-\frac{\alpha}{1-y}+\frac{\beta}{1+y}\right)-\frac{2}{\lambda^2\xi^2}(V-E)\right]P_n^{(\mu,\nu)} \\
&= \mathcal{A}_n(1-y)^\alpha(1+y)^\beta\frac{\xi^2}{1-y^2}\left\{(1-y^2)\frac{d^2}{dy^2}+\left[\frac{\xi'}{\xi}(1-y^2)+2(\beta-\alpha)-2y(\beta+\alpha)\right]\frac{d}{dy}\right. \\
&\left.+\frac{\xi'}{\xi}\left[\beta-\alpha-y(\beta+\alpha)\right]+\alpha(\alpha-1)\frac{1+y}{1-y}+\beta(\beta-1)\frac{1-y}{1+y}-2\alpha\beta-\frac{2(1-y^2)}{\lambda^2\xi^2}(V-E)\right\}P_n^{(\mu,\nu)}
\end{aligned}
\tag{B1}
$$

where $\xi(y)=\lambda^{-1}\frac{dy}{dx}$ and the prime stands for the derivative with respect to $y$. Using the differential equation of the Jacobi polynomial, $(1-y^2)\frac{d^2}{dy^2}P_n^{(\mu,\nu)}(y)=\left[(\mu+\nu+2)y+\mu-\nu\right]\frac{d}{dy}P_n^{(\mu,\nu)}(y)-n(n+\mu+\nu+1)P_n^{(\mu,\nu)}(y)$, we can rewrite (B1) as

$$
\begin{aligned}
\frac{-2}{\lambda^2}(H-E)\phi_n &= \mathcal{A}_n(1-y)^\alpha(1+y)^\beta\frac{\xi^2}{1-y^2}\left\{\left[\frac{\xi'}{\xi}(1-y^2)+(2\beta-\nu-2\alpha+\mu)-y(2\beta-\nu+2\alpha-\mu-2)\right]\frac{d}{dy}\right. \\
&\left.+\frac{\xi'}{\xi}\left[\beta-\alpha-y(\beta+\alpha)\right]+\alpha(\alpha-1)\frac{1+y}{1-y}+\beta(\beta-1)\frac{1-y}{1+y}-\left[2\alpha\beta+n(n+\mu+\nu+1)\right]-\frac{2(1-y^2)}{\lambda^2\xi^2}(V-E)\right\}P_n^{(\mu,\nu)}
\end{aligned}
\tag{B2}
$$

The tridiagonal requirement and the differential relation of the Jacobi polynomial, $(1-y^2)\frac{d}{dy}P_n^{(\mu,\nu)}(y)=-n\left(y+\frac{\nu-\mu}{2n+\mu+\nu}\right)P_n^{(\mu,\nu)}(y)+2\frac{(n+\mu)(n+\nu)}{2n+\mu+\nu}P_{n-1}^{(\mu,\nu)}(y)$, dictate that $(1-y^2)\frac{\xi'}{\xi}$ be linear in $y$. Thus, we may write $(1-y^2)\frac{\xi'}{\xi}=-a(1+y)+b(1-y)$ giving $\xi(y)=(1-y)^a(1+y)^b$ where $a$ and $b$ are real constants. Substituting in Eq. (B2) yields

$$
\begin{aligned}
-\frac{2}{\lambda^2}(H-E)\phi_n &= \mathcal{A}_n(1-y)^\alpha(1+y)^\beta\frac{\xi^2}{1-y^2}\left\{\left[(2\beta+b-\nu-2\alpha-a+\mu)-y(2\beta+b-\nu+2\alpha+a-\mu-2)\right]\frac{d}{dy}\right. \\
&\left.+\frac{2\alpha(\alpha+a-1)}{1-y}+\frac{2\beta(\beta+b-1)}{1+y}-\left[(\alpha+\beta)(\alpha+\beta+a+b-1)+n(n+\mu+\nu+1)\right]-\frac{2(1-y^2)}{\lambda^2\xi^2}(V-E)\right\}P_n^{(\mu,\nu)}
\end{aligned}
\tag{B3}
$$

Employing the differential relation of the Jacobi polynomial then multiplying from left by $\phi_m$ and integrating over $x$ while noting that $\lambda\int...dx=\int...\xi^{-1}dy$, we obtain the following matrix elements of the wave operator in the basis (43):

$$
\begin{aligned}
-\frac{2}{\lambda^2}\langle\phi_m|(H-E)|\phi_n\rangle &= \mathcal{A}_n\mathcal{A}_m\int_0^\infty(1-y)^\mu(1-y)^\nu\left[(1-y)^{2\alpha+a-\mu-1}(1+y)^{2\beta+b-\nu-1}\right]\times \\
&P_m^{(\mu,\nu)}(y)\left\{\left[\frac{2\beta+b-\nu-1}{1+y}-\frac{2\alpha+a-\mu-1}{1-y}\right]\left[-n\left(y+\frac{\nu-\mu}{2n+\mu+\nu}\right)P_n^{(\mu,\nu)}(y)+2\frac{(n+\mu)(n+\nu)}{2n+\mu+\nu}P_{n-1}^{(\mu,\nu)}(y)\right]\right. \\
&\left.+\left[\frac{2\alpha(\alpha+a-1)}{1-y}+\frac{2\beta(\beta+b-1)}{1+y}-\left[(\alpha+\beta)(\alpha+\beta+a+b-1)+n(n+\mu+\nu+1)\right]-\frac{2(1-y^2)}{\lambda^2\xi^2}(V-E)\right]P_n^{(\mu,\nu)}(y)\right\}dy
\end{aligned}
\tag{B4}
$$

Writing $2\alpha+a-\mu-1=p$ and $2\beta+b-\nu-1=q$, where $p$ and $q$ are integers, we can rewrite (B4) as follows

$$\frac{2}{\lambda^2}\langle\phi_m|(H-E)|\phi_n\rangle = n\langle m|(1-y)^p(1+y)^q\left[p+q-\frac{2p}{1-y}\frac{n+\nu}{2n+\mu+\nu}-\frac{2q}{1+y}\frac{n+\mu}{2n+\mu+\nu}\right]|n\rangle$$

$$-\frac{1}{2}\langle m|(1-y)^p(1+y)^q\left[\frac{(\mu+p)^2-(1-a)^2}{1-y}+\frac{(\nu+q)^2-(1-b)^2}{1+y}\right]|n\rangle \qquad \text{(B5)}$$

$$+\frac{1}{4}\langle m|(1-y)^p(1+y)^q\left[(2n+\mu+\nu+1)^2+(p+q+1)^2-(a+b-1)^2+2(\mu+\nu)(p+q)-1\right]|n\rangle$$

$$+(2n+\mu+\nu+1)D_{n-1}\langle m|(1-y)^p(1+y)^q\left(\tfrac{p}{1-y}-\tfrac{q}{1+y}\right)|n-1\rangle+\frac{2}{\lambda^2}\langle m|(1-y)^{p+1-2a}(1+y)^{q+1-2b}(V-E)|n\rangle$$

where we have defined $\langle y|n\rangle = \mathcal{A}_n(1-y)^{\mu/2}(1+y)^{\nu/2}P_n^{(\mu,\nu)}(y)$ and where $D_n$ is the recursion coefficient defined below Eq. (47). The orthogonality of the Jacobi polynomials, $\mathcal{A}_n^2\int_0^\infty(1-y)^\mu(1-y)^\nu P_n^{(\mu,\nu)}(y)P_m^{(\mu,\nu)}(y)dy = \delta_{n,m}$, and its recursion relation show that a tridiagonal matrix representation of the wave operator is obtained only in one of three cases:

(1) $(p,q)=(0,0)$, and $2\frac{1-y^2}{\lambda^2\xi^2}(V-E)=\frac{\mu^2-(1-a)^2}{2(1-y)}+\frac{\nu^2-(1-b)^2}{2(1+y)}+u_0+u_1y$, or (B6a)

(2) $(p,q)=(1,0)$, and $2\frac{1-y^2}{\lambda^2\xi^2}(V-E)=\frac{\nu^2-(1-b)^2}{2(1+y)}+\frac{u_-}{1-y}+u_0$, (B6b)

(3) $(p,q)=(0,1)$, and $2\frac{1-y^2}{\lambda^2\xi^2}(V-E)=\frac{\mu^2-(1-a)^2}{2(1-y)}+\frac{u_+}{1+y}+u_0$, or (B6c)

where $u_\pm$, $u_0$ and $u_1$ are arbitrary dimensionless parameters. Substituting these in (B5) and using the orthogonality of the Jacobi polynomials, we obtain

$$\frac{2}{\lambda^2}\langle\phi_m|(H-E)|\phi_n\rangle = \tfrac{1}{4}\left[(2n+\mu+\nu+1)^2-(a+b-1)^2+4u_0\right]\delta_{m,n}+u_1\langle m|y|n\rangle \qquad \text{(B7a)}$$

$$\frac{2}{\lambda^2}\langle\phi_m|(H-E)|\phi_n\rangle = \left[-\frac{2n(n+\nu)}{2n+\mu+\nu}-\frac{\mu+a}{2}(\mu-a+2)+u_-\right]\delta_{m,n}$$

$$+(2n+\mu+\nu+1)D_{n-1}\delta_{m,n-1}+\left[\left(n+\tfrac{\mu+\nu}{2}+1\right)^2-\tfrac{1}{4}(a+b-1)^2+u_0\right]\left(\delta_{m,n}-\langle m|y|n\rangle\right) \qquad \text{(B7b)}$$

$$\frac{2}{\lambda^2}\langle\phi_m|(H-E)|\phi_n\rangle = \left[-\frac{2n(n+\mu)}{2n+\mu+\nu}-\frac{\nu+b}{2}(\nu-b+2)+u_+\right]\delta_{m,n}$$

$$-(2n+\mu+\nu+1)D_{n-1}\delta_{m,n-1}+\left[\left(n+\tfrac{\mu+\nu}{2}+1\right)^2-\tfrac{1}{4}(a+b-1)^2+u_0\right]\left(\delta_{m,n}+\langle m|y|n\rangle\right) \qquad \text{(B7c)}$$

where $\langle n|y|m\rangle$ is the tridiagonal symmetric matrix *K* given by Eq. (47) (the Jacobi quadrature matrix).

# Appendix C
# Coordinate transformation in the Laguerre and Jacobi bases

In the Laguerre basis, the coordinate transformation $x\to y(x)$ is such that $y(x)\ge 0$ and $\frac{dy}{dx}=\lambda y^a e^{by}$. Integration yields $\lambda x=\int y^{-a}e^{-by}dy$, which is a match with the integral representation of the incomplete gamma function (see Sec. 6.5 of Ref. [20]). Thus, we obtain two results depending on the physical space (values of the parameter *a* and *b*):

$$\int_0^y t^{-a}e^{-bt}dt = b^{a-1}\gamma(1-a,by)\,, \tag{C1}$$

where $\gamma(x,y)$ is the lower incomplete gamma function, which is equal to $x^{-1}y^x\,{}_1F_1\left(\begin{smallmatrix}x\\x+1\end{smallmatrix}\middle|-y\right)$. Thus, we obtain

$$\lambda x = b^{a-1}\gamma(1-a,by) = \frac{y^{1-a}}{1-a}\,{}_1F_1\left(\begin{smallmatrix}1-a\\2-a\end{smallmatrix}\middle|-by\right),\; a\neq 1\,. \tag{C2}$$

On the other hand, changing the integration limits gives

$$\int_y^\infty t^{-a}e^{-bt}dt = b^{a-1}\Gamma(1-a,by)\,, \tag{C3}$$

where $\Gamma(x,y)$ is the upper incomplete gamma function, which is related to $\gamma(x,y)$ as $\Gamma(x,y)+\gamma(x,y)=\Gamma(x)$. Therefore, we obtain the alternative result

$$\lambda x = b^{a-1}\Gamma(1-a,by) = b^{a-1}\Gamma(1-a) - \frac{y^{1-a}}{1-a}\,{}_1F_1\left(\begin{smallmatrix}1-a\\2-a\end{smallmatrix}\middle|-by\right),\; a\neq 1\,. \tag{C4}$$

On the other hand, in the Jacobi basis the coordinate transformation $x\to y(x)$ is such that $-1\le y(x)\le +1$ and $\frac{dy}{dx}=\lambda(1-y)^a(1+y)^b$. Integration yields $\lambda x = \int(1-y)^{-a}(1+y)^{-b}dy$, which is a match with the integral representation of the incomplete beta function (see Sec. 6.6 of Ref. [20]) Thus, we obtain two results depending on the physical space (values of the parameter *a* and *b*):

$$\int_{-1}^y (1-t)^{-a}(1+t)^{-b}dt = 2^{1-a-b}B\left(\tfrac{1+y}{2};1-b,1-a\right), \tag{C5}$$

where $B(z;\alpha,\beta)$ is the incomplete beta function, which is equal to $\left(z^\alpha/\alpha\right){}_2F_1\left(\begin{smallmatrix}\alpha,1-\beta\\1+\alpha\end{smallmatrix}\middle|z\right)$. Therefore, we obtain (for $b\neq 1$)

$$\lambda x = \int_{-1}^y (1-t)^{-a}(1+t)^{-b}dt = 2^{1-a-b}B\left(\tfrac{1+y}{2};1-b,1-a\right) = \frac{(1+y)^{1-b}}{2^a(1-b)}\,{}_2F_1\left(\begin{smallmatrix}a,1-b\\2-b\end{smallmatrix}\middle|\tfrac{1+y}{2}\right), \tag{C6}$$

On the other hand, changing the integration limits and after some simple manipulations, we obtain the alternative result

$$\int_y^{+1} (1-t)^{-a}(1+t)^{-b}dt = 2^{1-a-b}B\left(\tfrac{1-y}{2};1-a,1-b\right), \tag{C7}$$

where $B\left(\tfrac{1-y}{2};\alpha,\beta\right)+B\left(\tfrac{1+y}{2};\beta,\alpha\right)=B(\alpha,\beta)=\frac{\Gamma(\alpha)\Gamma(\beta)}{\Gamma(\alpha+\beta)}$ is the complete beta function. Therefore, we obtain the alternative result for $a\neq 1$

$$\lambda x = \int_y^{+1} (1-t)^{-a}(1+t)^{-b}dt = 2^{1-a-b}B\left(\tfrac{1-y}{2};1-a,1-b\right) = \frac{(1-y)^{1-a}}{2^b(1-a)}\,{}_2F_1\left(\begin{smallmatrix}b,1-a\\2-a\end{smallmatrix}\middle|\tfrac{1-y}{2}\right). \tag{C8}$$

# Appendix D
# Orthogonal polynomials in the Laguerre basis

In the Laguerre bases, the matrix wave equation leads to three-term recursion relations for the expansion coefficients of the continuum wavefunction. These are identified with the three-term recursion relations of either the Meixner-Pollaczek polynomial or the continuous dual Hahn polynomial. For ease of reference, we give relevant properties of these two orthogonal polynomials and their discrete versions, which are used as expansion coefficients for the bound states wavefunction. Here, we consider the normalized version of these polynomials where $\int_{z_-}^{z_+}\rho(z)P_n(z)P_m(z)dz=\delta_{n,m}$, $P_0(z)=1$ and $\rho(z)$ is the normalized weight function. The orthogonality of the discrete version reads $\sum_{k_-}^{k_+}\omega(k)\,Q_n(k)Q_m(k)=\delta_{n,m}$, where $Q_n(k)$ is the normalized discrete version of $P_n(z)$ and $\omega(k)$ is the corresponding discrete weight function.

The normalized version of the Meixner-Pollaczek polynomial is (Sec. 1.7 of Ref. [6])

$$P_n^{\mu}(z,\theta)=\sqrt{\frac{\Gamma(n+2\mu)}{\Gamma(2\mu)\Gamma(n+1)}}\,e^{in\theta}\,{}_2F_1\left(\begin{smallmatrix}-n,\mu+iz\\2\mu\end{smallmatrix}\middle|1-e^{-2i\theta}\right), \tag{D1}$$

where $z\in[-\infty,+\infty]$, $\mu>0$ and $0<\theta<\pi$. This polynomial has only a continuous spectrum. It satisfies the following symmetric three-term recursion

$$\begin{aligned}\left(z\sin\theta\right)P_n^{\mu}(z,\theta)=&-\left[\left(n+\mu\right)\cos\theta\right]P_n^{\mu}(z,\theta)\\&+\frac{1}{2}\sqrt{n(n+2\mu-1)}P_{n-1}^{\mu}(z,\theta)+\frac{1}{2}\sqrt{(n+1)(n+2\mu)}P_{n+1}^{\mu}(z,\theta)\end{aligned} \tag{D2}$$

The corresponding normalized weight function is

$$\rho^{\mu}(z,\theta)=\frac{1}{2\pi\Gamma(2\mu)}(2\sin\theta)^{2\mu}e^{(2\theta-\pi)z}\left|\Gamma(\mu+iz)\right|^2. \tag{D3}$$

The asymptotics ($n\to\infty$) of this polynomial reads as follows (see, for example, the Appendix in Ref. [1])

$$P_n^{\mu}(z;\theta)\approx\frac{2n^{-1/2}e^{(\pi/2-\theta)z}}{(2\sin\theta)^{\mu}\left|\Gamma(\mu+iz)\right|}\cos\left[n\theta+\arg\Gamma(\mu+iz)-\mu\pi/2-z\ln(2n\sin\theta)\right] \tag{D4}$$

Therefore, ignoring the logarithmic term in the argument of the cosine relative to the linear term for large *n*, the scattering phase shift (mod $\pi/2$) reads as follows

$$\delta^{\mu}(z)=\arg\Gamma(\mu+iz)-\mu\tfrac{\pi}{2}. \tag{D5}$$

The scattering amplitude in (D4) vanishes if $\mu+iz=-n$ giving the discrete spectrum formula associated with this polynomial as $z^2=-\left(n+\mu\right)^2$. The spectrum is infinite for positive $\mu$ whereas it is finite for negative $\mu$ with $n=0,1,..,N$ and *N* is the largest integer less than or equal to $-\mu$. The generating function is (Sec. 1.7 of Ref. [6])

$$\sum_{n=0}^{\infty}\tilde{P}_n^{\mu}(z,\theta)t^n=\left(1-te^{i\theta}\right)^{-\mu+iz}\left(1-te^{-i\theta}\right)^{-\mu-iz}, \tag{D6}$$

where $\tilde{P}_n^{\mu}(z;\theta)=\sqrt{\frac{\Gamma(n+2\mu)}{\Gamma(2\mu)\Gamma(n+1)}}P_n^{\mu}(z;\theta)$. The discrete version of the polynomial is the Meixner (Krawtchouk) polynomial for infinite (finite) spectrum. The normalized version of the Meixner polynomial is written as (Sec. 1.9 of Ref. [6])

$$M_n^{\mu}\left(m;\beta\right)=\sqrt{\frac{\Gamma(n+2\mu)\,\beta^n}{\Gamma(2\mu)\Gamma(n+1)}}\,{}_2F_1\left(\begin{smallmatrix}-n,-m\\2\mu\end{smallmatrix}\middle|1-\beta^{-1}\right), \tag{D7}$$

where $0<\beta<1$ and $\mu>0$. It satisfies the following recursion relation

$$\begin{aligned}(1-\beta)m\,M_n^{\mu}\left(m;\beta\right)=&\left[n(1+\beta)+2\mu\beta\right]M_n^{\mu}\left(m;\beta\right)\\&-\sqrt{n(n+2\mu-1)\beta}M_{n-1}^{\mu}\left(m;\beta\right)-\sqrt{(n+1)(n+2\mu)\beta}M_{n+1}^{\mu}\left(m;\beta\right)\end{aligned} \tag{D8}$$

The normalized discrete weight function is $\rho_m^{\mu}(\beta)=\beta^m(1-\beta)^{2\mu}\frac{\Gamma(m+2\mu)}{\Gamma(2\mu)\Gamma(m+1)}$. On the other hand, the normalized version of the Krawtchouk polynomial is written as (Sec. 1.10 of Ref. [6])

$$K_n^N(m;\beta)=\sqrt{\tfrac{\Gamma(N+1)\beta^n/(1-\beta)^n}{\Gamma(N-n+1)\Gamma(n+1)}}\,{}_2F_1\left(\begin{smallmatrix}-n,-m\\-N\end{smallmatrix}\middle|\beta^{-1}\right), \tag{D9}$$

where $0<\beta<1$ and $n,m=0,1,2,..,N$ . It satisfies the following recursion relation

$$\begin{aligned} m K_n^N(m;\beta)&=\left[n(1-\beta)+\beta(N-n)\right]K_n^N(m;\beta)\\ &-\sqrt{\beta(1-\beta)n(N-n+1)}K_{n-1}^N(m;\beta)-\sqrt{\beta(1-\beta)(n+1)(N-n)}K_{n+1}^N(m;\beta)\end{aligned} \tag{D10}$$

The normalized discrete weight function is $\rho_m^N(\beta)=\beta^m(1-\beta)^{N-m}\frac{\Gamma(N+1)}{\Gamma(m+1)\Gamma(N-m+1)}$.

The normalized version of the continuous dual Hahn polynomial is (Sec. 1.3 of Ref. [6])

$$S_n^\mu(z^2;\alpha,\beta)=\sqrt{\tfrac{(\mu+\alpha)_n(\mu+\beta)_n}{n!(\alpha+\beta)_n}}\,{}_3F_2\left(\begin{smallmatrix}-n,\mu+\mathrm{i}z,\mu-\mathrm{i}z\\ \mu+\alpha,\mu+\beta\end{smallmatrix}\middle|1\right), \tag{D11}$$

where ${}_3F_2\left(\begin{smallmatrix}a,b,c\\ d,e\end{smallmatrix}\middle|z\right)=\sum_{n=0}^{\infty}\frac{(a)_n(b)_n(c)_n}{(d)_n(e)_n}\frac{z^n}{n!}$ is the generalized hypergeometric function and $(c)_n=c(c+1)(c+2)...(c+n-1)=\frac{\Gamma(n+c)}{\Gamma(c)}$. For a certain range of values of its parameters $\{\mu,\alpha,\beta\}$, this polynomial could have a continuous as well as a discrete spectrum. If the parameters are positive except for a pair of complex conjugates with positive real parts then the spectrum is purely continuous. However, if $\mu<0$ and $\mu+\alpha$, $\mu+\beta$ are positive or complex conjugates with positive real part then the spectrum is a mix of a continuous part and a discrete part. The polynomial satisfies the following symmetric three-term recursion relation

$$\begin{aligned} z^2S_n^\mu&=\left[(n+\mu+\alpha)(n+\mu+\beta)+n(n+\alpha+\beta-1)-\mu^2\right]S_n^\mu\\ &-\sqrt{n(n+\alpha+\beta-1)(n+\mu+\alpha-1)(n+\mu+\beta-1)}\,S_{n-1}^\mu\\ &-\sqrt{(n+1)(n+\alpha+\beta)(n+\mu+\alpha)(n+\mu+\beta)}\,S_{n+1}^\mu\end{aligned} \tag{D12}$$

The corresponding normalized weight function is

$$\rho^\mu(z;\alpha,\beta)=\frac{1}{2\pi}\frac{\left|\Gamma(\mu+\mathrm{i}z)\Gamma(\alpha+\mathrm{i}z)\Gamma(\beta+\mathrm{i}z)/\Gamma(2\mathrm{i}z)\right|^2}{\Gamma(\mu+\alpha)\Gamma(\mu+\beta)\Gamma(\alpha+\beta)}. \tag{D13}$$

The asymptotics ($n\to\infty$) of this polynomial reads as follows (see, for example, the Appendix in Ref. [1])

$$\begin{aligned} S_n^\mu(z^2;\alpha,\beta)&\approx\frac{2\sqrt{\Gamma(\mu+\alpha)\Gamma(\mu+\beta)\Gamma(\alpha+\beta)}\left|\Gamma(2\mathrm{i}z)\right|}{\left|\Gamma(\alpha+\mathrm{i}z)\Gamma(\beta+\mathrm{i}z)\Gamma(\mu+\mathrm{i}z)\right|\sqrt{n}}\times\\ &\cos\left\{z\ln n+\arg\left[\Gamma(2\mathrm{i}z)/\Gamma(\mu+\mathrm{i}z)\Gamma(\alpha+\mathrm{i}z)\Gamma(\beta+\mathrm{i}z)\right]\right\}\end{aligned} \tag{D14}$$

Since $\ln n\approx o(n^\xi)$ for any $\xi>0$, then we extract the following phase shift

$$\delta^\mu(z)=\arg\left[\Gamma(2\mathrm{i}z)/\Gamma(\mu+\mathrm{i}z)\Gamma(\alpha+\mathrm{i}z)\Gamma(\beta+\mathrm{i}z)\right]. \tag{D15}$$

The scattering amplitude in (D14) vanishes if $\mu+\mathrm{i}z=-n$ giving the associated (finite) discrete spectrum formula is $z^2=-(n+\mu)^2$, where $n=0,1,2,..,N$ and $N$ is the largest integer less than or equal to $-\mu$. The generating function is (Sec. 1.3 of Ref. [6])

$$\sum_{n=0}^{\infty}\tilde{S}_n^\mu(z^2;\alpha,\beta)t^n=(1-t)^{-\mu+\mathrm{i}z}\,{}_2F_1\left(\begin{smallmatrix}\alpha+\mathrm{i}z,\beta+\mathrm{i}z\\ \alpha+\beta\end{smallmatrix}\middle|t\right), \tag{D16}$$

where $\tilde{S}_n^\mu(z^2;\alpha,\beta)=\frac{(\mu+\alpha)_n(\mu+\beta)_n}{n!(\alpha+\beta)_n}\,{}_3F_2\left(\begin{smallmatrix}-n,\mu+\mathrm{i}z,\mu-\mathrm{i}z\\ \mu+\alpha,\mu+\beta\end{smallmatrix}\middle|1\right)$. The discrete version of the polynomial is the dual Hahn polynomial, which we write as (Sec. 1.6 of Ref. [6])

$$R_n^N(m;\alpha,\beta)=\sqrt{\frac{(\alpha+1)_n(\beta+1)_{N-n}}{n!(N-n)!}}\,{}_3F_2\left(\left.\begin{matrix}-n,-m,m+\alpha+\beta+1\\ \alpha+1,-N\end{matrix}\right|1\right), \tag{D17}$$

where $n,m=0,1,2,..,N$ and either $\alpha,\beta>-1$ or $\alpha,\beta<-N$ . It satisfies the following recursion relation

$$\begin{aligned}\left(m+\tfrac{\alpha+\beta+1}{2}\right)^2 R_n^N(m;\alpha,\beta) &= -\Big\{\left(n+\tfrac{\alpha+1}{2}\right)^2+\left(n-\tfrac{\beta+1}{2}\right)^2-N\left(2n+\alpha+1\right)\\ &\quad -\tfrac{1}{4}\Big[\left(\alpha+\beta+1\right)^2+\left(\alpha+1\right)^2+\left(\beta+1\right)^2\Big]\Big\}R_n^N(m;\alpha,\beta)\\ &\quad +\sqrt{n(n+\alpha)(N-n+1)(N-n+\beta+1)}R_{n-1}^N(m;\alpha,\beta)\\ &\quad +\sqrt{(n+1)(n+\alpha+1)(N-n)(N-n+\beta)}R_{n+1}^N(m;\alpha,\beta)\end{aligned} \tag{D18}$$

The normalized discrete weight function is

$$\rho^N(m;\alpha,\beta)=\left(N!\right)\frac{(2m+\alpha+\beta+1)(\alpha+1)_m(N-m+1)_m}{(m+\alpha+\beta+1)_{N+1}(\beta+1)_m m!}. \tag{D19}$$

Now, the overall asymptotics of the wavefunction should include not just those of the energy polynomials but also those of the basis elements of Eq. (2). Using the well-known asymptotics of the Laguerre polynomials, we obtain the following

$$\phi_n(y)\approx y^{\frac{1}{2}(2\alpha-\nu-1/2)}e^{-(\beta-1/2)y}\frac{1}{\sqrt{\pi\sqrt{n}}}\cos\left[2\sqrt{ny}-\tfrac{\pi}{2}\left(\nu+\tfrac{1}{2}\right)\right]. \tag{D20}$$

Consequently, this will contribute only a constant shift of $+\frac{\pi}{2}\left(\nu+\frac{1}{2}\right)$ to the scattering phase shift obtained above in (D5) and (D15). Note that the basis parameter $\nu$ for the scattering states of all physical systems presented in section 2 is energy independent. This supports our assertion that the basis functions do not carry information about the structure or dynamics of the physical system.

# Appendix E
# Orthogonal polynomials in the Jacobi basis

In the Jacobi bases, the matrix wave equation leads to three-term recursion relations for the expansion coefficients of the continuum wavefunction. However, these recursion relations and their associated orthogonal polynomials are not treated in the mathematic literature and their analytic properties (weight function, generating function, orthogonality, differential property, spectrum formula, asymptotics, etc.) are yet to be derived. Nonetheless, special cases of these polynomials have already been encountered in physics (for example, in the solution of the problem of an electron in the field of a molecule with an electric dipole moment [13]). In this Appendix, we define these orthogonal polynomials by their three-term recursion relations that enables us to obtain all of them analytically to any desired order starting with their normalized zero order. Moreover, we obtain the spectrum formula for special cases of these polynomials using the exact (finite/infinite) energy spectrum of the associated exactly solvable physical problems that belong to the conventional class. We identify here two orthogonal polynomials associated with the three cases (44a), (44b) and (44c).

The first orthogonal polynomial is associated with the Jacobi basis for the case (44a). It is a three-parameter polynomial designated as $H_n^{(\mu,\nu)}(z;\sigma)$. It satisfies the following symmetric three-term recursion relation (for $n=1,2,...$):

$$z\,H_n^{(\mu,\nu)}(z;\sigma)=\left(B_n^2+\sigma C_n\right)H_n^{(\mu,\nu)}(z;\sigma)+\sigma\left[D_{n-1}H_{n-1}^{(\mu,\nu)}(z;\sigma)+D_nH_{n+1}^{(\mu,\nu)}(z;\sigma)\right] \quad \text{(E1)}$$

where $B_n=n+\frac{\mu+\nu+1}{2}$, $C_n$ and $D_n$ are defined below Eq. (47). This recursion relation could be rewritten in an alternative (standard) format for the polynomial $\bar{H}_n^{(\mu,\nu)}(z;\sigma)$ as

$$\begin{aligned}z\,\bar{H}_n^{(\mu,\nu)}&=\left[\left(n+\frac{\mu+\nu+1}{2}\right)^2+\frac{\sigma(\nu^2-\mu^2)}{(2n+\mu+\nu)(2n+\mu+\nu+2)}\right]\bar{H}_n^{(\mu,\nu)}\\&+\frac{2\sigma(n+\mu)(n+\nu)}{(2n+\mu+\nu)(2n+\mu+\nu+1)}\bar{H}_{n-1}^{(\mu,\nu)}+\frac{2\sigma(n+1)(n+\mu+\nu+1)}{(2n+\mu+\nu+1)(2n+\mu+\nu+2)}\bar{H}_{n+1}^{(\mu,\nu)}\end{aligned} \quad \text{(E2)}$$

where $H_n^{(\mu,\nu)}(z;\sigma)$ is the normalized version of $\bar{H}_n^{(\mu,\nu)}(z;\sigma)$, which is defined by $H_n^{(\mu,\nu)}(z;\sigma)=\mathcal{A}_n\bar{H}_n^{(\mu,\nu)}(z;\sigma)$ with $\mathcal{A}_n=\sqrt{\left(2n+\mu+\nu+1\right)\frac{\Gamma(n+1)\Gamma(n+\mu+\nu+1)}{\Gamma(n+\mu+1)\Gamma(n+\nu+1)}}$. The polynomial solution of the recursion relation (E1) with $H_0^{(\mu,\nu)}(z)=1$ and $H_1^{(\mu,\nu)}(z)=\left(\sigma D_0\right)^{-1}\left[z-\sigma C_0-\frac{1}{4}\left(\mu+\nu+1\right)^2\right]$ is referred to as polynomial of the first kind. The polynomial of the second kind is obtained from the same recursion relation (E1) with $H_0^{(\mu,\nu)}=1$ but $H_1^{(\mu,\nu)}$ has one or two different linearity coefficients. Although that this polynomial is not treated in the mathematics literature, it was found in the solution of many problems in theoretical physics [12-16] where it was named the "*dipole polynomial*" [13]. The Jacobi polynomial is a special case of this polynomial, which is obtained as $P_n^{(\mu,\nu)}(y)=\lim_{\sigma,z\to\infty}\sigma\bar{H}_0^{(\mu,\nu)}(z;\sigma)$ such that $\lim_{\sigma,z\to\infty}\left(z/\sigma\right)=y\in[-1,+1]$. From the physics of the problems associated with this polynomial, we expect that it has only a continuous spectrum. In other words, its orthogonality is an integral over a continuous interval, which is infinite. The well-known energy spectra of the physical systems associated with the special cases of this polynomial presented in section 3.1 (including the trigonometric Scarf, hyperbolic Eckart, hyperbolic Pöschl-Teller and hyperbolic Rosen-Morse potentials) give the spectrum formula for this polynomial (with $\sigma=0$) as $z_n=\left(n+\frac{\mu+\nu+1}{2}\right)^2$. Moreover, these associated physical problems suggest that this polynomial has two discrete versions, one with an infinite spectrum and another that has a finite spectrum. We symbolize the former by $h_n^{(\mu,\nu)}(z_m;\sigma)$ and the latter by $k_n^{(\mu,\nu)}(z_m;\sigma)$ with the choice being dependent on the range of values of the parameters $\mu$ and $\nu$. These polynomials constitute the expansion coefficients of the $m^{\text{th}}$ bound state wavefunction. Additionally, the exact scattering phase shifts of the physical systems associated with $H_n^{(\mu,\nu)}(z;\sigma)$ could have also been used to give the phase shift in the asymptotics ($n\to\infty$) of this polynomial.

The second orthogonal polynomial is associated with the Jacobi basis for the case (44b). It is also a three-parameter polynomial, which is symbolized as $G_n^{(\mu,\nu)}(z;\sigma)$. It satisfies the following alternative three-term recursion relation

$$z\,G_n^{(\mu,\nu)}(z;\sigma)=\left[\frac{n(n+\nu)}{n+\frac{\mu+\nu}{2}}+\tfrac{1}{2}(\mu+1)^2+\left(\sigma+\tilde{B}_n^2\right)\left(C_n-1\right)\right]G_n^{(\mu,\nu)}(z;\sigma)$$
$$+\left(\sigma+\tilde{B}_{n-1}^2\right)D_{n-1}\,G_{n-1}^{(\mu,\nu)}(z;\sigma)+\left(\sigma+\tilde{B}_n^2\right)D_n\,G_{n+1}^{(\mu,\nu)}(z;\sigma) \qquad \text{(E3)}$$

where $\tilde{B}_n=n+\frac{\mu+\nu}{2}+1=B_n+\frac{1}{2}$, $C_n$ and $D_n$ are defined below Eq. (47). Similarly, this recursion relation could be rewritten in the following alternative format for the polynomial $\bar{G}_n^{(\mu,\nu)}(z;\sigma)$ as

$$z\,\bar{G}_n^{(\mu,\nu)}=\left\{\frac{2n(n+\nu)}{2n+\mu+\nu}+\tfrac{1}{2}(\mu+1)^2+\left(\sigma+\tilde{B}_n^2\right)\left[\frac{\nu^2-\mu^2}{(2n+\mu+\nu)(2n+\mu+\nu+2)}-1\right]\right\}\bar{G}_n^{(\mu,\nu)}$$
$$+\left(\sigma+\tilde{B}_{n-1}^2\right)\frac{2(n+\mu)(n+\nu)}{(2n+\mu+\nu)(2n+\mu+\nu+1)}\bar{G}_{n-1}^{(\mu,\nu)}+\left(\sigma+\tilde{B}_n^2\right)\frac{2(n+1)(n+\mu+\nu+1)}{(2n+\mu+\nu+1)(2n+\mu+\nu+2)}\bar{G}_{n+1}^{(\mu,\nu)} \qquad \text{(E4)}$$

where $G_n^{(\mu,\nu)}(z;\sigma)$ is the normalized version of $\bar{G}_n^{(\mu,\nu)}(z;\sigma)$, which is defined by $G_n^{(\mu,\nu)}(z;\sigma)=\mathcal{A}_n\bar{G}_n^{(\mu,\nu)}(z;\sigma)$. Again, the polynomial of the first kind satisfies (E3) for $n=1,2,...$ with $G_0^{(\mu,\nu)}=1$ and $G_1^{(\mu,\nu)}=D_0^{-1}\left\{1-C_0+\left(\sigma+\tilde{B}_0^2\right)^{-1}\left[z-\frac{1}{2}(\mu+1)^2\right]\right\}$. Although this polynomial is not treated in the mathematics literature, it was encountered in physics while searching for a solution of the Schrödinger equation with the S-wave Hulthén potential [12]. From the physics of the problems associated with this polynomial, we expect that it has either a continuous spectrum or a mix of continuous and discrete spectra. This depends on the range of values of its parameters $\{\mu,\nu,\sigma\}$. Moreover, the discrete spectrum is finite. Figure 7 is an illustration of the mix spectrum case. It shows the distribution of the zeros of the polynomial $G_N^{(\mu,\nu)}(z;\sigma)$ on the *z*-axis for a large order *N* and for a given set of values of its parameters. It is obvious that for such values the spectrum is a mix. Now, the well-known energy spectra of the problems associated with the special cases of this polynomial presented in section 3.2 (including the trigonometric Scarf, hyperbolic Eckart and hyperbolic Pöschl-Teller potentials) give the spectrum formula for this polynomial (with $\sigma<0$) as $z_n=2\left(n+\frac{\nu+1}{2}-\sqrt{-\sigma}\right)^2$ where $n=0,1,..,N$ and *N* is the largest integer less than or equal to $\sqrt{-\sigma}-\frac{\nu+1}{2}$ and $\mu$ depends linearly on the physical parameters $\{\sigma,\nu,N\}$. For completely confined systems where there are only bound states, the expansion coefficients of the $m^{\text{th}}$ bound state wavefunction will be the discrete version of the polynomial $G_n^{(\mu,\nu)}(z;\sigma)$ whose spectrum is finite and which we denote as $g_n^{(\mu,\nu)}(z_m;\sigma)$. Moreover, the exact scattering phase shifts associated with these potentials do also give the phase shift in the asymptotics ($n\to\infty$) of the polynomial $G_n^{(\mu,\nu)}(z;\sigma)$.

Now, the overall asymptotics of the wavefunction should include not just those of the energy polynomials (which are not yet known, unfortunately) but also those of the basis elements of Eq. (43). Using the well-known asymptotics of the Jacobi polynomials, we obtain the following

$$\phi_n(y)\approx\sqrt{\frac{2}{\pi}}\,(1-y)^{\frac{1}{2}(2\alpha-\mu-1/2)}(1+y)^{\frac{1}{2}(2\beta-\nu-1/2)}\cos\left[\theta\left(n+\frac{\mu+\nu+1}{2}\right)-\frac{\pi}{2}\left(\mu+\frac{1}{2}\right)\right], \qquad \text{(E5)}$$

where $\cos\theta = y$. Consequently, this will contribute only a constant shift of $+\frac{\pi}{2}\left(\mu+\frac{1}{2}\right)$ to the scattering phase shift obtained from the energy polynomials $H_n^{(\mu,\nu)}(z;\sigma)$ and $G_N^{(\mu,\nu)}(z;\sigma)$. Here too, the basis parameter $\mu$ for the scattering states of all physical systems presented in section 3 is energy independent. This again supports our assertion that the basis functions do not carry information about the structure or dynamics of the physical system.

# Appendix F
# Non-tridiagonal overlap matrix Ω in the Jacobi basis

For special cases in the Jacobi class of problems, we could maintain the tridiagonal structure of the wave operator $\mathcal{J} = H - E\Omega$ even if Ω is not. Since the potential is energy independent, then a counter term from the kinetic energy part of the Hamiltonian must be chosen to eliminate it as demonstrated below. Using the symmetry mentioned in the first paragraph of section 3, we treat here cases that correspond to (*a*,*b*) overlooking those that correspond to (*b*,*a*). For example, if we treat the case $(a,b)=\left(1,\frac{1}{2}\right)$ then we do not consider the case $(a,b)=\left(\frac{1}{2},1\right)$, which is easily obtained from the former by a simple parameter map and $y \to -y$. Here, we consider only the class of problems with the most general physical configuration corresponding to Eq. (44a) where $\Omega_{n,m} = \langle n|(1-y)^{1-2a}(1+y)^{1-2b}|m\rangle$ and $V(x)-E=(1-y)^{2a-1}(1+y)^{2b-1}\left[\frac{U_+}{1+y}+\frac{U_-}{1-y}+U_0+U_1 y\right]$ where $U_+=\frac{\lambda^2}{4}\left[\nu^2-(1-b)^2\right]$ and $U_-=\frac{\lambda^2}{4}\left[\mu^2-(1-a)^2\right]$. Now, if the right-hand side of the equation for $V(x)-E$ has a non-zero constant term, say $U_{const}$, then we can always eliminate the non-tridiagonal Ω matrix by choosing $E=V_{const}-U_{const}$, where $V_{const}$ is the constant part of the potential. We select cases where $V_{const}\neq U_{const}$ so that the solution is valid for all energies not just $E=0$. However, for demonstration purposes we present below (in F.4) one case where the solution is valid only for $E=0$. Now, the condition $V_{const}\neq U_{const}$ is satisfied only if physical requirements place a restriction on the value of one or more potential parameters. An example of such physical requirements is the vanishing of the potential function at infinity.

**F.1: $(a,b)=(1,0)$:**

$\Omega_{n,m}=\langle n|\frac{1+y}{1-y}|m\rangle$, $y(x)=1-2e^{-\lambda x}$ with $x\geq 0$, $U_+=\frac{\lambda^2}{4}(\nu^2-1)$, $U_-=\frac{\lambda^2}{4}\mu^2$ and

$$\begin{aligned} V-E &= \frac{1-y}{1+y}\left[\frac{U_+}{1+y}+\frac{U_-}{1-y}+U_0+U_1 y\right] \\ &= U_+\frac{1-y}{(1+y)^2}+\frac{U_-+2(U_0-U_1)}{1+y}-U_1 y+2U_1-U_0 \end{aligned} \tag{F1}$$

where we have used $\frac{1-y}{1+y}=-1+\frac{2}{1+y}$ and $\frac{y}{1+y}=1-\frac{1}{1+y}$. We also write the potential as

$$V(y)=\frac{1-y}{1+y}\left[\frac{V_+}{1+y}+\frac{V_-}{1-y}+V_0+V_1y\right], \tag{F2}$$

and to force it to vanish at infinity ( $y=1$ ), we require that $V_-=0$. Substituting (F2) in the left side of (F1), we obtain finally

$$\nu^2=1+\tfrac{4}{\lambda^2}V_+,\; U_1=V_1,\; U_0=V_0+E \text{ and } \mu^2=-8E/\lambda^2 . \tag{F3}$$

**F.2: $(a,b)=\left(1,\tfrac{1}{2}\right)$:**

$\Omega_{n,m}=\langle n|(1-y)^{-1}|m\rangle$, $y(x)=2\tanh^2(\lambda x)-1$ with $x\geq 0$ and $\lambda\to\lambda\sqrt{2}$, $U_+=\tfrac{\lambda^2}{2}\times\left(\nu^2-\tfrac{1}{4}\right)$, $U_-=\tfrac{\lambda^2}{2}\mu^2$ and

$$\begin{aligned}V-E&=(1-y)\left[\frac{U_+}{1+y}+\frac{U_-}{1-y}+U_0+U_1y\right]\\&=\frac{2U_+}{1+y}+(U_1-U_0)y-U_1y^2-U_++U_-+U_0\end{aligned} \tag{F4}$$

where, again, we used $\tfrac{1-y}{1+y}=-1+\tfrac{2}{1+y}$. We also write the potential as

$$V(y)=(1-y)\left[\frac{V_+}{1+y}+\frac{V_-}{1-y}+V_0+V_1y\right], \tag{F5}$$

and to force the potential to vanish at infinity ( $y=1$ ), we require that $V_-=0$. Substituting (F5) in the left side of (F4), we obtain finally

$$\nu^2=\tfrac{1}{4}+\tfrac{2}{\lambda^2}V_+,\; U_1=V_1,\; U_0=V_0 \text{ and } \mu^2=-2E/\lambda^2 . \tag{F6}$$

**F.3: $(a,b)=(1,1)$:**

$\Omega_{n,m}=\langle n|\left(1-y^2\right)^{-1}|m\rangle$, $y(x)=\tanh(\lambda x)$ with $-\infty<x<+\infty$, $U_+=\tfrac{\lambda^2}{4}\nu^2$, $U_-=\tfrac{\lambda^2}{4}\mu^2$ and

$$\begin{aligned}V-E&=\left(1-y^2\right)\left[\frac{U_+}{1+y}+\frac{U_-}{1-y}+U_0+U_1y\right]\\&=U_++U_-+(U_--U_+)y+\left(1-y^2\right)(U_0+U_1y)\end{aligned} \tag{F7}$$

We also write the potential as

$$V(y)=\left(1-y^2\right)\left[\frac{V_+}{1+y}+\frac{V_-}{1-y}+V_0+V_1y\right], \tag{F8}$$

and to force it to vanish at infinity ( $y=\pm 1$ ), we require that $V_\pm=0$. Substituting (F8) into the left side of (F7) with $V_\pm=0$, we obtain finally

$$U_1=V_1,\; U_0=V_0 \text{ and } \mu^2=\nu^2=-2E/\lambda^2 . \tag{F9}$$

**F.4: $(a,b)=\left(\tfrac{3}{2},\tfrac{1}{2}\right)$:**

$\Omega_{n,m}=\langle n|(1-y)^{-2}|m\rangle$, $y(x)=\dfrac{(\lambda x)^2-1}{(\lambda x)^2+1}$ for $x\geq 0$, $U_+=\dfrac{\lambda^2}{4}\left(\nu^2-\tfrac{1}{4}\right)$, $U_-=\dfrac{\lambda^2}{4}\left(\mu^2-\tfrac{1}{4}\right)$

and

$$V-E=(1-y)^2\left[\frac{U_+}{1+y}+\frac{U_-}{1-y}+U_0+U_1y\right]$$
$$=\frac{4U_+}{1+y}+\left(U_+-U_--2U_0+U_1\right)y+\left(U_0-2U_1\right)y^2+U_1y^3-3U_++U_-+U_0 \tag{F10}$$

where, again, we used $\frac{1-y}{1+y}=-1+\frac{2}{1+y}$ and $\frac{y}{1+y}=1-\frac{1}{1+y}$. We write the potential as

$$V(y)=(1-y)^2\left[\frac{V_+}{1+y}+\frac{V_-}{1-y}+V_0+V_1y\right]. \tag{F11}$$

Substituting (F11) in the left side of (F10), we obtain $U_{const}=V_{const}$. Thus, we should take $E=0$, which means that an exact solution is obtained only at zero energy. If $E\neq 0$ then we have to resort to numerical routines to get a solution as was done for this particular potential in subsection 3.1.6.

## Table Captions:

**Table 1**: Lowest part of the energy spectrum (in units of $-\frac{1}{2}\lambda^2$) for the power-law radial potential (22) and for different sizes of the basis set. We took $u_0=-7$, $u_1=2$ and $\ell=1$. $\varepsilon=2E/\lambda^2$ and $u_i=2V_i/\lambda^2$ for all tables.

**Table 2**: Lowest part of the energy spectrum (in units of $\frac{1}{2}\lambda^2$) for the logarithmic potential box (26) and for different sizes of the basis set. We took $u_0=1$, $u_1=-5$ and $u_2=2$.

**Table 3**: Lowest part of the energy spectrum (in units of $\frac{1}{2}\lambda^2$) for the potential box (41) and for different sizes of the basis set. We took $\nu=\frac{1}{2}$, $u_1=-1$ and $u_2=1$. Slower convergence is evident for this potential.

**Table 4**: Lowest part of the energy spectrum (in units of $\frac{1}{2}\lambda^2$) for the potential box (50) and for different sizes of the basis set. We took $u_0=0$, $u_1=-3$, $u_+=1$ and $u_-=2$.

**Table 5**: Lowest part of the energy spectrum (in units of $\frac{1}{2}\lambda^2$) for the potential box (50) with $u_0=0$, $u_\pm=0$, $u_1=5$ and for a basis size of 20. Our results are in excellent agreement with those of Table I in Ref. [14].

**Table 6**: Lowest part of the energy spectrum (in units of $\frac{1}{2}\lambda^2$) for the potential box (53) and for different sizes of the basis set. We took $u_0=-3$, $u_1=5$, $u_+=1$ and $u_-=2$.

**Table 7**: The finite energy spectrum (in units of $-\frac{1}{2}\lambda^2$) for the potential (55) with $u_0=-50$, $u_1=10$ and $u_+=5$. It was obtained by diagonalizing the Hamiltonian matrix for different sizes of the basis set.

**Table 8**: The finite energy spectrum (in units of $-\frac{1}{2}\lambda^2$) for the potential (59) with $u_0=-70$, $u_1=10$ and $u_+=5$. It was obtained by diagonalization of the Hamiltonian matrix for different sizes of the basis set.

**Table 9**: The finite energy spectrum (in units of $-\frac{1}{2}\lambda^2$) for the potential (65) with $u_0=-30$ and $u_1=20$. It was obtained by diagonalization of the Hamiltonian matrix for different sizes of the basis set.

**Table 10**: The finite energy spectrum (in units of $-\frac{1}{2}\lambda^2$) for the potential function (69) and for different sizes of the basis set. We took $u_0=-50$, $u_1=30$, $u_+=2$ and $u_-=1$.

**Table 11**: Lowest part of the energy spectrum (in units of $\frac{1}{2}\lambda^2$) for the potential box (71) and for different sizes of the basis set. We took $u_0 = -5$, $u_1 = 3$, $u_+ = 2$ and $u_- = 1$.

**Table 12**: List of solvable potential in the TRA that do not belong to the conventional class of exactly solvable problems. These are generalized versions of known potentials except for the new potential box in the second row of the Table. Generalization is accomplished by adding a new term (the $V_1$ term). The two potentials in the second and third row were never studied in the published literature. This list does not exhaust the class of potentials that are exactly solvable in the TRA. See, for example, the potential function (84).

## Figure Captions:

**Fig. 1**: Plot of $y(x)$ obtained by inverting the coordinate transformation $x(y)$ of Eq. (40) with $0 \le x \le L$ and $\lambda = \sqrt{2\pi}/L$.

**Fig. 2**: Plot of the 1D potential box given by Eq. (41). The graph is obtained by varying $u_2$ from –5 (bottom trace) to +5 (top trace) in steps of +2 while keeping $u_1 = 2$. We define $u_i = 2V_i/\lambda^2$.

**Fig. 3**: Plot of the potential box given by Eq. (53) with $\lambda = 2\sqrt{2}/L$. Part (a) is obtained by varying $u_1$ from –5 (bottom right side trace) to +5 (top right side trace) in steps of +2 while keeping $u_0 = -3$, $u_+ = 2$ and $u_- = 1$. Part (b) is obtained by varying $u_+$ from 0 (bottom trace) to 5 (top trace) in steps of +1 while keeping $u_0 = -2$, $u_1 = 3$ and $u_- = 1$.

**Fig. 4**: Plot of the potential given by Eq. (69). Part (a) is obtained by varying $u_1$ from –5 (top left side trace) to +5 (bottom left side trace) in steps of +2 while keeping $u_0 = -1$, $u_+ = 1$ and $u_- = 0$. We define $u_i = 2V_i/\lambda^2$. Part (b) is obtained by varying $u_+$ from 0 (bottom trace) to 5 (top trace) in steps of +1 while keeping $u_0 = -3$, $u_1 = -2$ and $u_- = 1$.

**Fig. 5**: Plot of $y(x)$ obtained by inverting the coordinate transformation $x(y)$ of Eq. (71) with $-L/2 \le x \le +L/2$ and $\lambda = \pi/2L$.

**Fig. 6**: Plot of the potential box given by Eq. (71). Part (a) is obtained by varying $u_1$ from –5 (bottom right side trace) to +5 (top right side trace) in steps of +2 while keeping the other potential parameters fixed at $u_0 = 1$, $u_+ = +2$ and $u_- = -1$. We define $u_i = 2V_i/\lambda^2$. Part (b) is obtained by varying $u_+$ from –1 (bottom trace) to +4 (top trace) in steps of +1 while fixing the others at $u_0 = -4$, $u_- = 2$ and $u_1 = 3$.

**Fig. 7**: The distribution of zeros of the polynomial $G_N^{(\mu,\nu)}(z;\sigma)$ along the z-axis for $\mu = 1$ , $\nu = 2$, $\sigma = -12$ and for large *N*. It is obvious that for such parameter values the spectrum consists of a continuous part on $z \le 0$ and a discrete part made up of two points at *z* = 7.71539 and *z* = 1.85898.

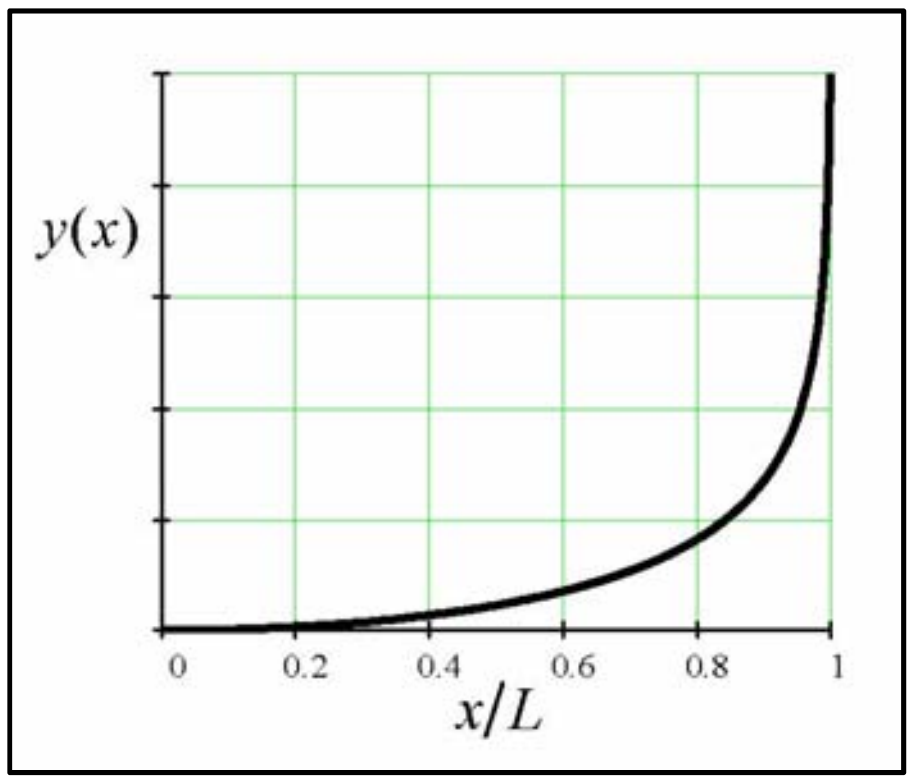

**Fig.1**

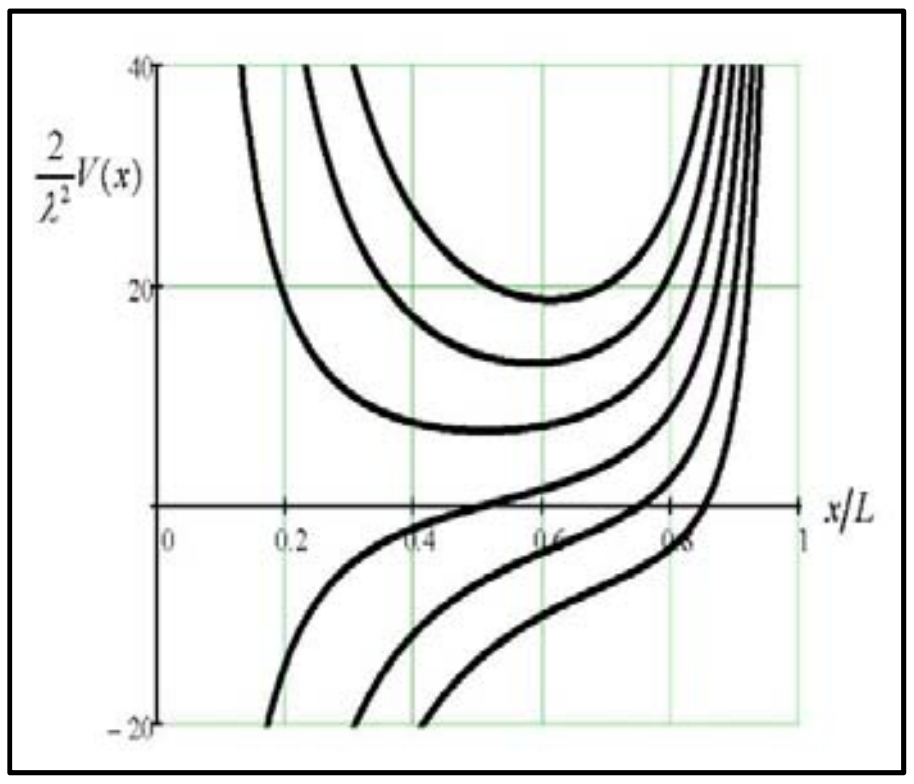

**Fig. 2**

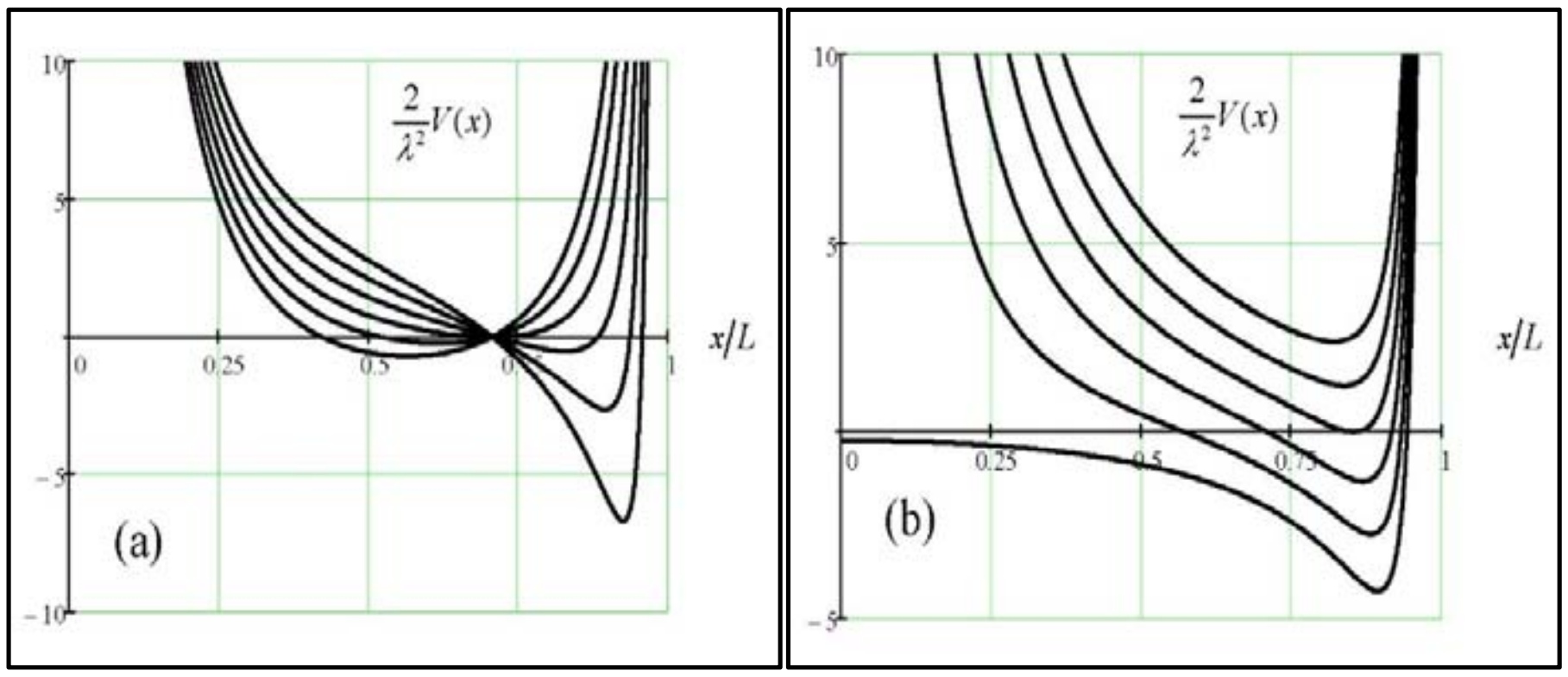

**Fig. 3**

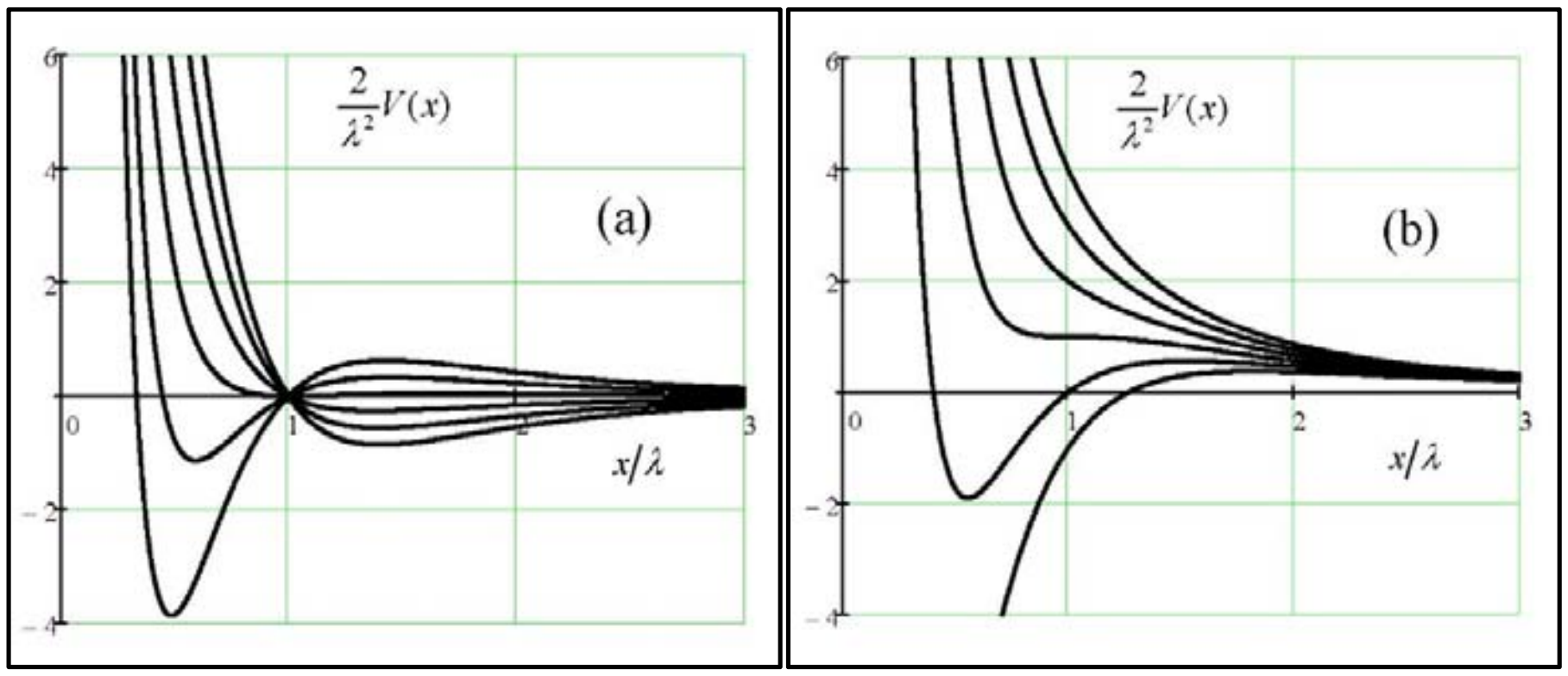

**Fig. 4**

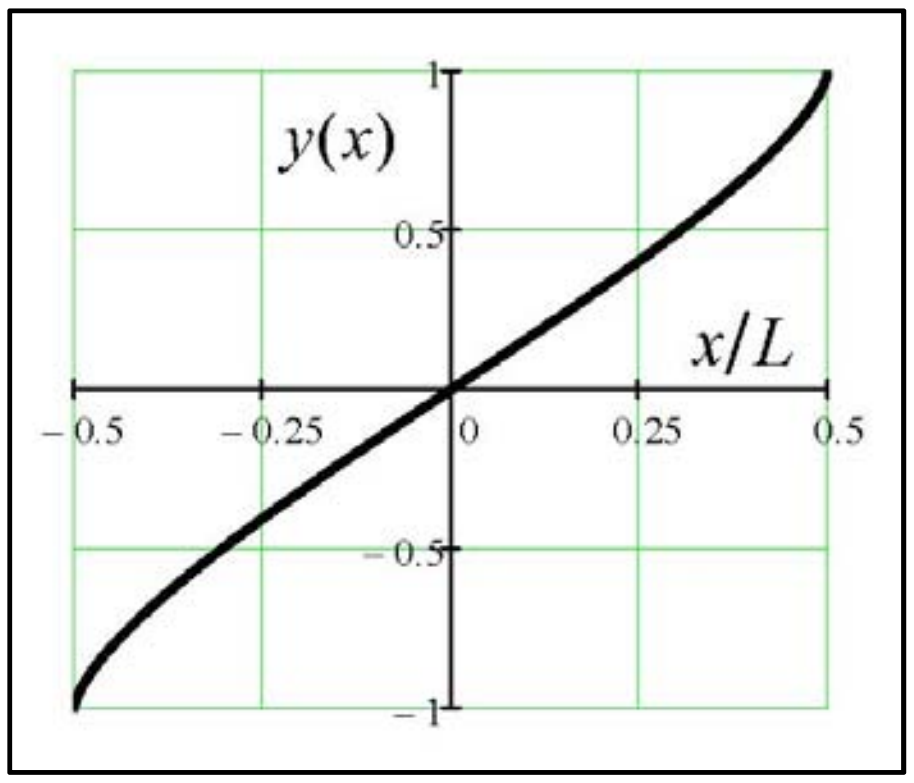

**Fig. 5**

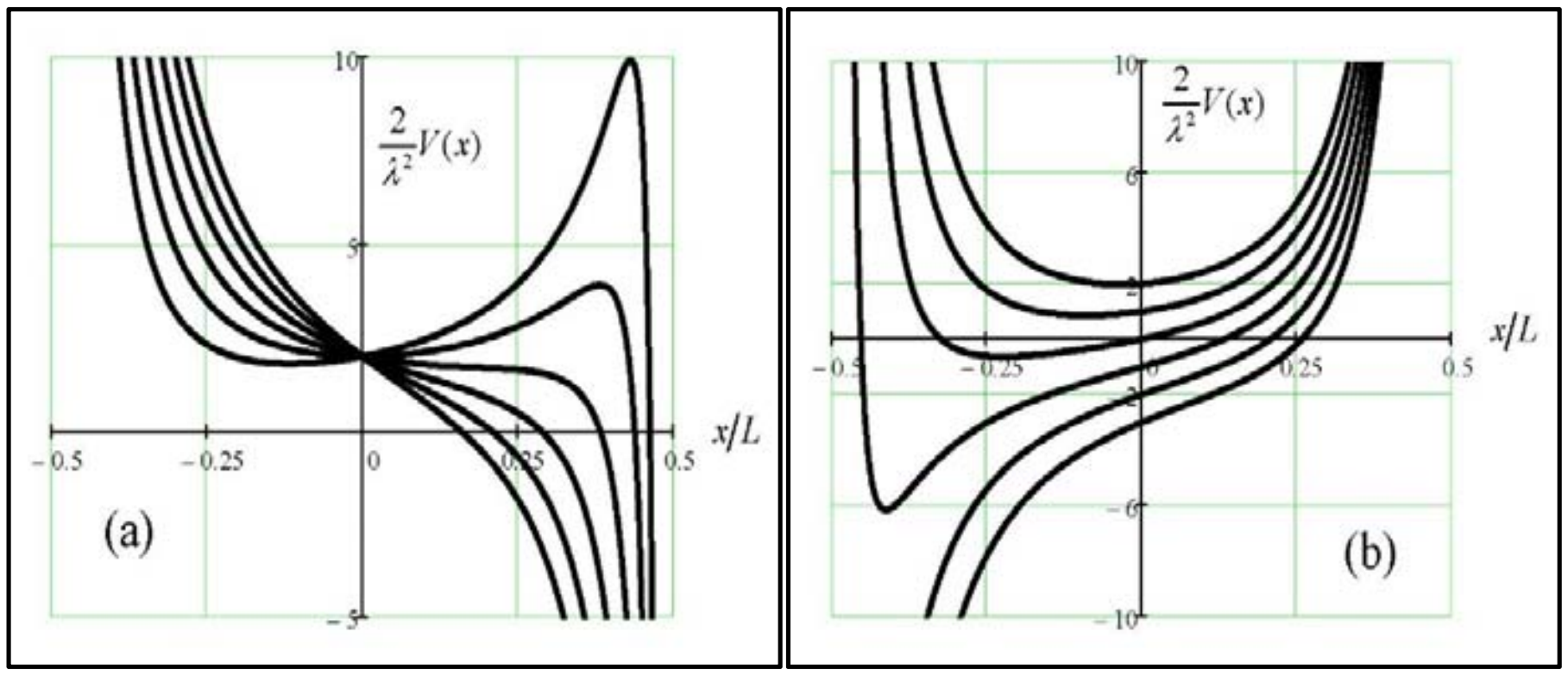

**Fig. 6**

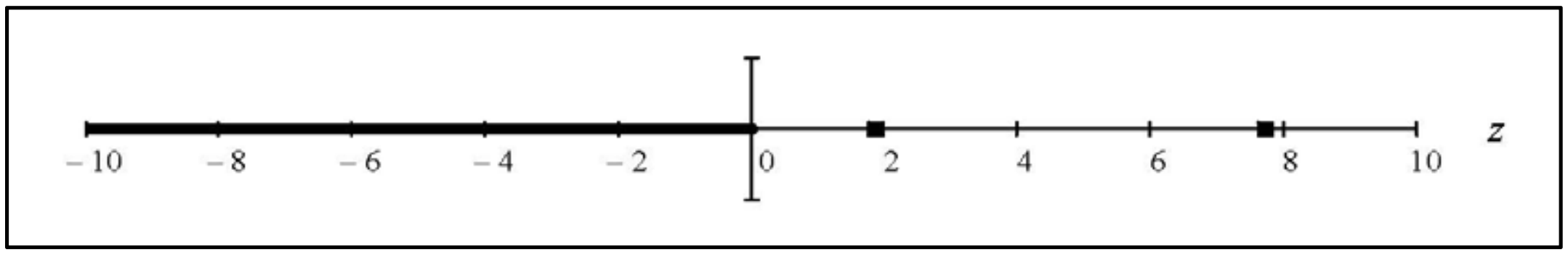

**Fig. 7**

**Table 1**

| *n* | 20×20 | 30×30 | 50×50 | 300×300 |
|---|---|---|---|---|
| 0 | 1.8298990 | 1.8297104 | 1.8297060 | 1.8297060 |
| 1 | 1.2396343 | 1.2381818 | 1.2381656 | 1.2381655 |
| 2 | 0.9414138 | 0.9367592 | 0.9367971 | 0.9367972 |
| 3 | 0.7651641 | 0.7536586 | 0.7539073 | 0.7539072 |
| 4 | 0.6458098 | 0.6302285 | 0.6310113 | 0.6310105 |
| 5 | 0.5473050 | 0.5409642 | 0.5427047 | 0.5427028 |
| 6 | 0.4628143 | 0.4743339 | 0.4761633 | 0.4761601 |
| 7 | 0.3928129 | 0.4297753 | 0.4242103 | 0.4242060 |
| 8 | 0.3354517 | 0.3954936 | 0.3825244 | 0.3825092 |
| 9 | 0.2882833 | 0.3545833 | 0.3483966 | 0.3483004 |

**Table 2**

| *n* | 30×30 | 50×50 | 100×100 | 300×300 |
|---|---|---|---|---|
| 0 | –3.45191 | –3.45191 | –3.45191 | –3.45191 |
| 1 | 6.32506 | 6.32511 | 6.32511 | 6.32511 |
| 2 | 21.66442 | 21.68532 | 21.68528 | 21.68528 |
| 3 | 43.45214 | 42.36163 | 42.34894 | 42.34894 |
| 4 | 61.81710 | 67.64252 | 68.18969 | 68.18967 |
| 5 | 110.03849 | 99.32792 | 99.14510 | 99.14042 |
| 6 | 321.17267 | 144.89319 | 135.02164 | 135.16038 |
| 7 | 1185.34101 | 178.02987 | 176.31113 | 176.22249 |
| 8 | 5338.67654 | 382.57012 | 224.90390 | 222.30773 |
| 9 | 29089.23294 | 960.45860 | 258.16798 | 273.40209 |

**Table 3**

| *n* | 30×30 | 50×50 | 100×100 | 300×300 | 500×500 |
|---|---|---|---|---|---|
| 0 | 3.4627 | 3.4616 | 3.4612 | 3.4610 | 3.4610 |
| 1 | 10.7087 | 10.6869 | 10.6797 | 10.6779 | 10.6777 |
| 2 | 21.2584 | 21.1242 | 21.0723 | 21.0615 | 21.0608 |
| 3 | 32.2583 | 34.9434 | 34.6677 | 34.6231 | 34.6205 |
| 4 | 53.2743 | 49.7428 | 51.5149 | 51.3591 | 51.3516 |
| 5 | 124.3649 | 68.0863 | 71.8121 | 71.2688 | 71.2499 |
| 6 | 363.3289 | 118.9417 | 94.1271 | 94.3569 | 94.3136 |
| 7 | 1291.4057 | 242.6353 | 113.8956 | 120.6361 | 120.5432 |
| 8 | 5540.4655 | 559.8656 | 156.6755 | 150.1344 | 149.9421 |
| 9 | 28643.2306 | 1448.5403 | 239.5919 | 182.9137 | 182.5172 |

**Table 4**

| *n* | 10×10 | 20×20 | 200×200 |
|---|---|---|---|
| 0 | 5.258544076432 | 5.258544076432 | 5.258544076432 |
| 1 | 10.916769371149 | 10.916769371149 | 10.916769371149 |
| 2 | 18.439002773109 | 18.439002773109 | 18.439002773109 |
| 3 | 27.968329593297 | 27.968329593297 | 27.968329593297 |
| 4 | 39.508546472695 | 39.508546472695 | 39.508546472695 |
| 5 | 53.055831996542 | 53.055831996542 | 53.055831996542 |
| 6 | 68.607516101327 | 68.607516101138 | 68.607516101138 |
| 7 | 86.162006491471 | 86.162006241114 | 86.162006241114 |
| 8 | 105.718578104034 | 105.718349478201 | 105.718349478200 |
| 9 | 127.369164902386 | 127.275958149897 | 127.275958149897 |

**Table 5**

| *n* | This work | Ref. [14] |
|---|---|---|
| 0 | –0.5955395589892 | –0.5955395589892 |
| 1 | 4.3453451696558 | 4.3453451696558 |
| 2 | 9.3549646941811 | 9.3549646941810 |
| 3 | 16.2001100732554 | 16.2001100732554 |
| 4 | 25.1266923657196 | 25.1266923657196 |
| 5 | 36.0875520021223 | 36.0875520021223 |
| 6 | 49.0641568653650 | 49.0641568653649 |
| 7 | 64.0490437059899 | 64.0490437059898 |
| 8 | 81.0387114884925 | 81.0387114884928 |
| 9 | 100.0313345578343 | 100.0313345578344 |

**Table 6**

| *n* | 20×20 | 50×50 | 200×200 |
|---|---|---|---|
| 0 | 0.972760968732 | 0.972760968732 | 0.972760968735 |
| 1 | 6.983408121093 | 6.983408121093 | 6.983408121097 |
| 2 | 15.373343726289 | 15.373343726289 | 15.373343726293 |
| 3 | 26.180137389613 | 26.180137389613 | 26.180137389628 |
| 4 | 39.421935063305 | 39.421935063305 | 39.421935063318 |
| 5 | 55.108718929868 | 55.108718929868 | 55.108718929868 |
| 6 | 73.246519168447 | 73.246519168447 | 73.246519168445 |
| 7 | 93.839240612875 | 93.839240612714 | 93.839240612712 |
| 8 | 116.889550235645 | 116.889550021682 | 116.889550021679 |
| 9 | 142.399427222307 | 142.399346572748 | 142.399346572746 |

**Table 7**

| $n$ | 20×20 | 50×50 | 100×100 | 200×200 |
|---|---|---|---|---|
| 0 | 147.816766580900 | 147.816766580902 | 147.816766580947 | 147.816766580928 |
| 1 | 50.012953295875 | 50.012953295876 | 50.012953295875 | 50.012953295873 |
| 2 | 17.662783992105 | 17.662783992106 | 17.662783992106 | 17.662783992105 |
| 3 | 5.337458183693 | 5.337458095523 | 5.337458095497 | 5.337458095496 |
| 4 | 0.879514930138 | 0.879468623113 | 0.879467000406 | 0.879466871701 |

**Table 8**

| $n$ | 20×20 | 50×50 | 100×100 | 300×300 |
|---|---|---|---|---|
| 0 | 80.730895189951 | 80.730895189951 | 80.730895189953 | 80.730895189970 |
| 1 | 46.216665984094 | 46.216665984087 | 46.216665984090 | 46.216665984094 |
| 2 | 21.626689530935 | 21.626689492479 | 21.626689492466 | 21.626689492466 |
| 3 | 6.510654288386 | 6.510534637678 | 6.510533362484 | 6.510533321607 |
| 4 | 0.407520316423 | 0.319311958059 | 0.290206704796 | 0.269685348895 |

**Table 9**

| $n$ | 20×20 | 50×50 | 100×100 | 300×300 |
|---|---|---|---|---|
| 0 | 27.093164564541 | 27.093164546466 | 27.093164546465 | 27.093164546467 |
| 1 | 16.852247030136 | 16.852246971225 | 16.852246971196 | 16.852246971191 |
| 2 | 9.038079680128 | 9.037956937932 | 9.037956483468 | 9.037956476279 |
| 3 | 3.665812482976 | 3.665729503567 | 3.665727299403 | 3.665727134971 |
| 4 | 0.825015093605 | 0.736473792738 | 0.719660752896 | 0.713081581957 |

**Table 10**

| $n$ | 20×20 | 50×50 | 100×100 |
|---|---|---|---|
| 0 | 163.9220892483 | 163.9220892483 | 163.9220892483 |
| 1 | 91.5800542367 | 91.5800542367 | 91.5800542367 |
| 2 | 41.5962877093 | 41.5962877093 | 41.5962877093 |
| 3 | 12.6644072112 | 12.6644072130 | 12.6644072130 |
| 4 | 1.3459089781 | 1.3459050340 | 1.3459050340 |

**Table 11**

| $n$ | 20×20 | 50×50 | 200×200 |
|---|---|---|---|
| 0 | 2.236938203767 | 2.236938203767 | 2.236938203769 |
| 1 | 15.123421228744 | 15.123421228743 | 15.123421228743 |
| 2 | 36.363285538440 | 36.363285538439 | 36.363285538440 |
| 3 | 65.773142802859 | 65.773142803828 | 65.773142803826 |
| 4 | 103.286953770566 | 103.286953272074 | 103.286953272073 |
| 5 | 148.872078219253 | 148.872156272303 | 148.872156272303 |
| 6 | 202.513685913996 | 202.509964616559 | 202.509964616563 |
| 7 | 264.082136795116 | 264.188428675896 | 264.188428675901 |
| 8 | 334.906660695398 | 333.899419922014 | 333.899419922020 |
| 9 | 403.094428098940 | 411.637129223234 | 411.637129223235 |

**Table 12**

| $y(x)$ | $V(x)$ | $\mu^2$ | $\nu^2$ | $2\alpha$ | $2\beta$ | Bound | Scattering |
|---|---|---|---|---|---|---|---|
| $\sin(\pi x/L)$ <br> $-L/2 \le x \le +L/2$ | $V_0 + \dfrac{W_+ - W_- \sin(\pi x/L)}{\cos^2(\pi x/L)} + V_1 \sin(\pi x/L)$ <br> $W_\pm = V_+ \pm V_-$, $V_\pm \ge -(\pi/4L)^2$ | $\frac{1}{4} + \frac{L^2 V_-}{\pi^2/4}$ | $\frac{1}{4} + \frac{L^2 V_+}{\pi^2/4}$ | $\mu + \frac{1}{2}$ | $\nu + \frac{1}{2}$ | Infinite | No |
| $2(x/L)^2 - 1$ <br> $0 \le x \le L$ | $\dfrac{1/4}{1-(x/L)^2}\left[2V_0 + \dfrac{V_+}{(x/L)^2} + \dfrac{V_-}{1-(x/L)^2}\right] - V_1 \dfrac{(x/L)^2 - \frac{1}{2}}{(x/L)^2 - 1}$ <br> $V_+ \ge -1/2L^2$, $V_- \ge -2/L^2$ | $1 + \frac{L^2 V_-}{2}$ | $\frac{1}{4} + \frac{L^2 V_+}{2}$ | $\mu + 1$ | $\nu + \frac{1}{2}$ | Infinite | No |
| $2\tanh^2(\lambda x) - 1$ <br> $x \ge 0$ | $\dfrac{V_+}{\sinh^2(\lambda x)} + 2\dfrac{V_0 + V_1[2\tanh^2(\lambda x) - 1]}{\cosh^2(\lambda x)}$ <br> $V_+ \ge -\lambda^2/8$, $V_- = 0$, $E \le 0$ | $-\frac{2E}{\lambda^2}$ | $\frac{1}{4} + \frac{2V_+}{\lambda^2}$ | $\mu$ | $\nu + \frac{1}{2}$ | Finite | Yes |
| $\tanh(\lambda x)$ <br> $-\infty < x < +\infty$ | $\dfrac{V_0 + V_1 \tanh(\lambda x)}{\cosh^2(\lambda x)}$ <br> $V_\pm = 0$, $E \le 0$ | $-\frac{2E}{\lambda^2}$ | $-\frac{2E}{\lambda^2}$ | $\mu$ | $\nu$ | Finite | Yes |
| $1 - 2e^{-\lambda x}$ <br> $x \ge 0$ | $\dfrac{1}{e^{\lambda x} - 1}\left[V_0 + V_1\left(1 - 2e^{-\lambda x}\right) + \dfrac{V_+/2}{1 - e^{-\lambda x}}\right]$ <br> $V_+ \ge -(\lambda/2)^2$, $V_- = 0$, $E \le 0$ | $-\frac{8E}{\lambda^2}$ | $1 + \frac{4V_+}{\lambda^2}$ | $\mu$ | $\nu + 1$ | Finite | Yes |